\documentclass[11pt,draft]{IEEEtran}

\usepackage{psfig}      
\usepackage{color}      
\usepackage{epsf,psfrag,amssymb,amsfonts,latexsym,cite,verbatim}

\def\psfancypar#1#2{\begingroup\def\par{\endgraf\endgroup\lineskiplimit=0pt}
               \setbox2=\hbox{\large\sc #2}
               \newdimen\tmpht \tmpht \ht2 \advance\tmpht by \baselineskip
               \font\hhuge=Times-Bold at \tmpht
               \setbox1=\hbox{{\hhuge #1}}
               \count7=\tmpht \count8=\ht1
               \divide\count8 by 1000 \divide\count7 by \count8 
               \tmpht=.001\tmpht\multiply\tmpht by \count7 
               \font\hhuge=Times-Bold at \tmpht
               \setbox1=\hbox{{\hhuge #1}}
               \noindent
                \hangindent1.05\wd1
               \hangafter=-2 {\hskip-\hangindent
               \lower1\ht1\hbox{\raise1.0\ht2\copy1}%
                \kern-0\wd1}\copy2\lineskiplimit=-1000pt}

\def\boxit#1{\vbox{\hrule\hbox{\vrule\kern3pt
        \vbox{\kern3pt#1\kern3pt}\kern3pt\vrule}\hrule}}

\def\reals{ { {\rm  I \kern-0.15em R }  } }
\def\complex{ {\,{{\rm C} \kern-0.50em \raise0.20ex {  |}}\, }}

\def\mubf{\hbox{\boldmath$\mu$\unboldmath}}

\def\Sigmabf{\hbox{$\bf \Sigma$}}

\def\Pibf{{\bf \Pi}}

\def\hbf{{\bf h}}

\def\sbf{{\bf s}}

\def\ubf{{\bf u}}

\def\xbf{{\bf x}}
\def\ybf{{\bf y}}

\def\xbf{{\bf x}}
\def\ybf{{\bf y}}
\def\Abf{{\bf A}}
\def\Bbf{{\bf B}}

\def\Ibf{{\bf I}}

\def\Kbf{{\bf K}}

\def\Pbf{{\bf P}}
\def\Qbf{{\bf Q}}
\def\Rbf{{\bf R}}

\def\Nc{{\cal N}}

\def\be{\vskip .3cm \begin{equation}}
\def\ee{\end{equation} \vskip .4cm \noindent}
\def\defeq{{\stackrel{\Delta}{=}}}
\def\Rxx{\Rbf_{\ssstyle X\kern-.1em X}}

\let\ssstyle=\scriptscriptstyle

\def\Kout{\setbox1=\hbox{\Huge\bf K}\hbox to
1.05\wd1{\hspace{.05\wd1}
}}
\def\Sout{\setbox1=\hbox{\Huge\bf S}\hbox to 1.05\wd1{\hspace{.05\wd1}
}}

\def\scalefig#1{\epsfxsize #1\textwidth}
\def\defeq{\stackrel{\Delta}{=}}
\def\nn{\nonumber}

\newcommand {\Ebb}{{\mathbb{E}}}

\newcommand{\beq}{\begin{equation}}
\newcommand{\eeq}{\end{equation}}

\newtheorem{theorem}{Theorem}
\newtheorem{lemma}{Lemma}

\title{{\Large\bf Neyman-Pearson Detection of Gauss-Markov Signals in Noise: Closed-Form Error Exponent and Properties}}

\author{Youngchul Sung, Lang Tong$^\dagger$\thanks{$^\dagger$Corresponding author.}, and H. Vincent Poor
\thanks{
    Youngchul Sung is with Qualcomm Inc., San Diego, CA 92121, USA, E-mail: ysung@qualcomm.com. Lang Tong
    is  with the School of Electrical and
    Computer Engineering, Cornell University, Ithaca, NY 14853. Tel/Fax:
    (607) 255-3900/9072. Email: ltong@ece.cornell.edu.
    H. Vincent Poor is with Department of Electrical Engineering, Princeton University, Princeton, NJ 08544.
    Email: poor@princeton.edu.
}
\thanks{This work was supported in part by
the Multidisciplinary University Research Initiative (MURI)  under
the Office of Naval Research Contract N00014-00-1-0564.  Prepared
through collaborative participation in the Communications and
Networks Consortium sponsored by the U. S. Army Research
Laboratory under the Collaborative Technology Alliance Program,
Cooperative Agreement DAAD19-01-2-0011. The work of H. V. Poor was
also supported in part by the Office of Naval Research under Grant
N00014-03-1-0102. } } 
\begin{document}
\maketitle

\def\asadd#1{{\textbf{\textsl #1}}}
\def\ignore#1{}

\begin{abstract}

The performance of Neyman-Pearson detection of correlated random
signals using noisy observations is considered.  Using the large
deviations principle, the performance is analyzed via the {\em
error exponent} for the miss probability with a fixed false-alarm
probability. Using the state-space structure of the signal and
observation model, a closed-form expression for the  error
exponent  is derived using the innovations approach, and  the
connection between the asymptotic behavior of the optimal detector
and that of the Kalman filter is established. The properties of
the error exponent are investigated for the scalar case. It is
shown that the error exponent has distinct characteristics with
respect to correlation strength: for signal-to-noise ratio (SNR)
$\ge 1$, the error exponent is monotonically decreasing as the
correlation becomes strong while for SNR $<1$ there is an optimal
correlation that maximizes the error exponent for a given SNR.

\end{abstract}

\section{Introduction}\label{sec:intro}

In this paper, we consider the detection of correlated random
signals using noisy observations $y_i$ under the Neyman-Pearson
formulation. The null and alternative hypotheses are given by
\begin{equation}  \label{eq:hypothesisscalarstate}
\begin{array}{lcl}
H_0 &: & y_i = w_i,  ~~~~i=1,2,\cdots, n,\\
H_1 &: & y_i = s_i+ w_i,\\
\end{array}
\end{equation}
where  $\{w_i\}$ is independent and identically distributed
(i.i.d.) $\Nc(0,\sigma^2)$ noise with a known variance $\sigma^2$,
and $\{s_i\}$ is the stochastic signal process correlated in time.
We assume that $\{s_i\}$ is a Gauss-Markov process following a
state-space model. An example of an application in which this type
of problem arises is  the detection of stochastic signals in large
sensor networks, where it is reasonable to assume that signal
samples taken at closely spaced locations are correlated,  while
the measurement noise is independent from sensor to sensor.
 In this paper, we are interested in the performance of the
Neyman-Pearson detector for the hypotheses
(\ref{eq:hypothesisscalarstate})  with a fixed level (i.e.,
upper-bound constraint on the false-alarm probability) when the
sample size $n$ is large.

In many cases, the miss probability $P_M$ of the Neyman-Pearson
detector with a fixed level decays exponentially as the sample
size increases, and the {\em error exponent} is defined as the
rate of exponential decay, i.e.,
\begin{equation}
K \defeq \lim_{n\rightarrow\infty} -\frac{1}{n} \log P_M
\end{equation}
under the given false-alarm constraint.
 The error exponent is an important
parameter  since it gives an estimate of the number of samples
required for a given detector performance; faster decay rate
implies that fewer samples are needed for a given miss
probability, or that better performance can be obtained with a
given number of samples. Hence, the error exponent is a good
performance index for detectors in the large sample regime. For
the case of i.i.d. samples where each sample is drawn
independently from the common null probability density $p_0$ or
alternative density $p_1$, the  error exponent under the fixed
false-alarm constraint is given by the Kullback-Leibler
information $D(p_0||p_1)$ between the two densities $p_0$ and
$p_1$ (C. Stein \cite{Bahadur:71SIAM}).  For more general cases,
the  error exponent is given by the asymptotic Kullback-Leibler
rate
 defined as the almost-sure limit of
 \begin{equation}  \label{eq:asymKLrate}
\frac{1}{n} \log \frac{p_{0,n}}{p_{1,n}}(y_1,\cdots,y_n)
~~~\mbox{as} ~n\rightarrow \infty,
 \end{equation}
 under $p_{0,n}$, where
$p_{0,n}$  and $p_{1,n}$ are the null  and alternative joint
densities of $y_1,\cdots,y_n$, respectively, assuming that the
limit exists\footnote{Ergodic cases are examples for which this
limit exists.}\cite{Bahadur&Zabell&Gupta:80book, Vajda:book,
Vajda:90SPA, Chen:96IT, Luschgy&Rukhin&Vajda:93SPA}. However, the
closed-form calculation of (\ref{eq:asymKLrate}) is available only
for restricted cases. One such example is the discrimination
between two autoregressive (AR) signals with distinct parameters
under the two  hypotheses
\cite{Luschgy&Rukhin&Vajda:93SPA,Luschgy:94SJS}.  In this case,
the joint density, $p_{j,n}$,  is easily decomposed using the
Markov property under each  hypothesis, and the calculation of the
rate is straightforward.   However, for the problem of
(\ref{eq:hypothesisscalarstate}) this approach is not available
since the observation samples under the alternative hypothesis do
not possess the Markov property due to the additive noise, even if
the signal itself is Markovian; i.e., the alternative is a hidden
Markov model.

\subsection{Summary of Results}

Our approach to this problem is to exploit  the state-space model.
The state-space approach in detection is well established in
calculation of the log-likelihood ratio (LLR) for  correlated
signals\cite{Schweppe:65IT,VanTrees:70PROC}. With the state-space
model, the LLR is expressed through the innovations
representation\cite{Kailath:70PROC} and the innovations are easily
obtained by the Kalman filter. The key idea for the closed-form
calculation of the error exponent for the hidden Markov model is
based on the properties of innovations. Since the innovations
process is independent from time to time, the joint density under
$H_1$ is given by the product of marginal densities of the
innovations, and the LLR is given by a function of the sum of
squares of the innovations; this functional form facilitates the
closed-form calculation of (\ref{eq:asymKLrate}).

By applying this state-space approach,
 we derive  a closed-form expression for the error exponent
 $K$  for the miss probability of the Neyman-Pearson detector for (\ref{eq:hypothesisscalarstate})
 of fixed false-alarm probability, $P_F =\alpha$.

We next investigate the properties of the error exponent using the
obtained  closed-form expression.  We explore the asymptotic
relationship between the innovations approach and the spectrum of
the observation. We show that the error exponent $K$ is a function
of the signal-to-noise ratio (SNR) and the correlation, and has
different behavior with respect to (w.r.t.) the correlation
strength depending on the SNR. We show a sharp phase transition at
SNR = 1: at high SNR, $K$ is monotonically decreasing  as a
function of the correlation, while at low SNR, on the other hand,
there exists an optimal correlation value that yields the maximal
$K$.

We also make a connection between the asymptotic behavior of the
Kalman filter and that of the Neyman-Pearson detector.  It  is shown
that the error exponent is determined by the asymptotic (or
steady-state) variances of the innovations under $H_0$ and $H_1$
together with the noise variance.

\subsection{Related Works}

The detection of Gauss-Markov processes in Gaussian noise is a
classical problem. See \cite{Kailath&Poor:98IT} and references
therein.   Our work focuses on the performance analysis as
measured by the error exponent, and relies on the connection
between the likelihood ratio and the innovations process as
described by Schweppe \cite{Schweppe:65IT}.  In addition to the
calculation of the LLR, the state-space approach has also been
used in the performance analysis in this detection problem.
Exploiting the state-space model, Schweppe obtained a differential
equation for the Bhattacharyya distance between two Gaussian
processes\cite{Schweppe:67IC2,Schweppe:67IC,Kailath:67IT}, which
gives an upper bound on the average error probability under a
Bayesian formulation.

There is an extensive literature on the large deviations approach
to the analysis of the detection of Gauss-Markov processes
\cite{Donsker&Varadhan:85CMP, Benitz&Bucklew:90IT, Bahr:90IT,
Bahr&Bucklew:90SP, Barone&Gigli&Piccioni:95IT,
Bryc&Smolenski:93SPL, Bryc&Dembo:97JTP,
Bercu&Gamboa&Rouault:97SPA, Bercu&Rouault:02TPA,
Chamberland&Veeravalli:04ITWS}. Many of these results  rely on the
extension of Cramer's theorem by G\"artner and Ellis
\cite{Hollander:book,Dembo&Zeitouni:book,Gartner:77TVP,Ellis:84AP}
and the properties of the asymptotic eigenvalue distributions of
Toeplitz matrices\cite{Grenander&Szego:book,Gray:72IT}.   To find
the rate function, however, this approach usually requires an
optimization that requires nontrivial numerical methods except in
some simple cases, and the rate is given as an integral of the
spectrum of the observation process; closed-form expressions are
difficult to obtain except for the case of a noiseless
autoregressive (AR) process in discrete-time and its
continuous-time counterpart, the Ornstein-Uhlenbeck
process\cite{Donsker&Varadhan:85CMP, Benitz&Bucklew:90IT,
Bahr:90IT, Bahr&Bucklew:90SP, Barone&Gigli&Piccioni:95IT,
Bryc&Smolenski:93SPL, Bryc&Dembo:97JTP,
Bercu&Gamboa&Rouault:97SPA, Bercu&Rouault:02TPA}. In addition,
most results have been obtained for a fixed threshold for the
normalized LLR test, which results in expressions for the rate as
a function of the threshold. For ergodic cases, however, the
normalized LLR converges to a constant under the null hypothesis
and the false alarm probability also decays exponentially for a
fixed threshold. Hence, a detector with a fixed threshold is not
optimal in the Neyman-Pearson sense since it does not use the
false-alarm constraint fully; i.e., the optimal threshold is a
function of sample size.

\subsection{Notation and Organization}     \label{sec:notation}

We will make use of standard notational conventions.
    Vectors and matrices are written
    in boldface with matrices in capitals. All vectors are column vectors.
For a scalar $z$, $z^*$ denotes the complex conjugate. For a
matrix $\Abf$, $\Abf^T$ and $\Abf^H$ indicate the transpose and
Hermitian transpose, respectively. $\mbox{det} (\Abf)$ and
$\mbox{tr} (\Abf)$ denote the determinant and trace of $\Abf$,
respectively. $\Abf(l,m)$ denotes the element of the $l$th row and
$m$th column, and $\{\lambda_k(\Abf)\}$ denotes the set of all
eigenvalues of $\Abf$. We reserve $\Ibf_m$ for
    the identity matrix of size $m$ (the subscript is included only
    when necessary).
  For a sequence of random vectors $\xbf_n$, $\Ebb_{j}\{\xbf_n \}$ is
  the expectation of $\xbf_n$ under probability density $p_{j,n}, ~j=0,1$.
    The notation $\xbf\sim \Nc(\mubf,\Sigmabf)$   means that $\xbf$
    has the multivariate Gaussian distribution  with mean $\mubf$ and
    covariance $\Sigmabf$.

The paper is organized as follows. The data model is described in
Section \ref{sec:datamodel}. In Section \ref{sec:errorexponentCF},
the closed-form error exponent is obtained via the innovations
approach  representation. The properties of the error exponent are
investigated in Section \ref{sec:errorexp_properties}, and the
extension to the vector case is provided in Section
\ref{sec:vectormodel}. Simulation results
 are presented to demonstrate the predicted behavior in Section \ref{sec:numerical},
 followed by the conclusion in Section \ref{sec:conclusion}.

\section{Data Model} \label{sec:datamodel}

For the purposes of exposition, we will focus primarily on the
case in which the signal is generated by a scalar time-invariant
state space model.  The more general vector case will be
considered below. In particular, we assume that the signal process
$\{s_i\}$ has a time-invariant state-space structure
\begin{eqnarray}
s_{i+1} &=& a s_i +  u_i, ~~i=1,\cdots,n, \label{eq:statespacescalar} \\
s_1 &\sim& \Nc(0,~\Pi_0), \nn\\
 u_i  &\stackrel{i.i.d.}{\sim}& \Nc(0, ~Q),~~~Q=\Pi_0 (1-a^2), \nn
\end{eqnarray}
where $a$ and $\Pi_0$  are known scalars with $0 \le a \le 1$ and
$\Pi_0 \ge  0$. We assume that  the process noise $\{u_i\}$ is
independent of the measurement noise $\{w_i\}$ and  the initial
state $s_1$ is independent of  $u_i$ for all $i$.   Notice that
the signal sequence $\{s_i\}$ forms a stationary  process for this
choice of $Q$. Due to this stationarity, the signal variance is
$\Pi_0$ for all $i$, and the SNR  $\Gamma$ for the observations is
thus given by
\begin{equation}
\Gamma = \frac{\Pi_0}{\sigma^2}. \label{eq:SNRDEFscalar}
\end{equation}
Notice that the value of $a$ determines
the amount of correlation between signal samples.
 For an i.i.d. signal we have  $a=0$ and all the signal power
 results from the process noise $\{u_i\}$.  When the signal is perfectly correlated
 on the other hand, $a=1$ and the signal depends only on the realization of
 the  initial state $s_1$.  The autocovariance function
 $r_s(\cdot)$ of the signal process $\{s_i\}$ is given by
\begin{equation} \label{eq:sigmabfsscalar}
r_s(i-j)\defeq \Ebb \{s_i s_j\}= \Pi_0 a^{|i-j|}.
\end{equation}

As seen in (\ref{eq:hypothesisscalarstate}), the observation $y_i$
under the alternative hypothesis is given by a sum of signal
sample $s_i$ and independent noise $w_i$.   Thus, the observation
sequence $\{y_i\}$ under $H_1$ is  not a Markov process
   due to the presence of the additive noise
   even if the signal is Markovian.
Let $r_y^{(j)}(\cdot)$ denote the autocovariance  function of the
observation process $\{y_i\}$ under $H_j$, i.e.,
\begin{equation}
 r_y^{(j)}(m-n)= \Ebb_j \{y_m y_n\},
\end{equation}
and let  $S_y^{(j)}(\omega)$ be the spectrum  of the observation
process under $H_j$, i.e.,
 \begin{equation}
\label{eq:spectraldensity}
 S_y^{(j)}(\omega)= \sum_{k=-\infty}^\infty
r_y^{(j)}(k)e^{-jk\omega}, ~~-\pi \le \omega \le \pi.
\end{equation} Then, the spectra of the observation process
under $H_0$ and $H_1$ are given by
\begin{equation}  \label{eq:spectrumunderH1}
S_y^{(0)}(\omega)=\sigma^2, ~~~S_y^{(1)}(\omega) = \sigma^2 +
S_s(\omega), ~~-\pi \le \omega \le \pi,
\end{equation}
 where the signal spectrum under the state-space model is given
by the  Poisson kernel:
\begin{equation}
S_s(\omega) = \frac{\Pi_0(1-a^2)}{1-2a \cos \omega + a^2}, ~~~ 0
\le a <1 \label{eq:spectraldensityGM}.
\end{equation}

\section{Error Exponent for Gauss-Markov Signal in Noise} \label{sec:errorexponentCF}

In this section, we derive the error exponent of the
Neyman-Pearson detector with a fixed level $\alpha \in (0, 1)$ for
the Gauss-Markov signal
 described by (\ref{eq:statespacescalar}) embedded in noisy observations.

A general approach to the error exponent of Neyman-Pearson
detection of Gaussian signals can be framed in the spectral
domain. It is well known that the Kullback-Leibler information
between two zero-mean Gaussian distributions $p_{0}=
\Nc(0,\sigma_0^2)$ and $p_{1} = \Nc(0,\sigma_1^2)$ is given by
\begin{equation}  \label{eq:GaussianKLscalar}
D(p_{0}||p_{1}) =  \frac{1}{2} \log
\frac{\sigma_{1}^2}{\sigma_0^2} + \frac{1}{2}
 \frac{\sigma_{0}^2}{\sigma_{1}^2}-\frac{1}{2}.
\end{equation}
As noted above, this quantity gives the error exponent in the case
of an i.i.d. Gaussian signal. In more general cases with
correlated Gaussian signals, the error exponent can similarly be
obtained using the asymptotic properties of covariance matrices.
Let $\ybf_n$ be the random vector of observation samples $y_i$
defined as
\begin{equation}
\ybf_n \defeq [y_1, y_2, \cdots, y_n]^T.
\end{equation}
For two distributions $p_{0,n}(\ybf_n)=\Nc({\mathbf 0},
\Sigmabf_{0,n})$ and $p_{1,n}(\ybf_n) = \Nc({\mathbf 0},
\Sigmabf_{1,n})$, the error exponent is given by the almost-sure
limit of the Kullback-Leibler rate:
\begin{equation}  \label{eq:GaussianKLvector}
\frac{1}{n}\log \frac{p_{0,n}}{p_{1,n}}(\ybf_n) =
\frac{1}{n}\left(\frac{1}{2} \log \left(\frac{\det
(\Sigmabf_{1,n})}{ \det (\Sigmabf_{0,n})}\right) + \frac{1}{2}
\ybf_n^T (\Sigmabf_{1,n}^{-1}-\Sigmabf_{0,n}^{-1})\ybf_n\right),
\end{equation}
under $p_{0,n}$ \cite{Bahadur&Zabell&Gupta:80book, Vajda:book,
Vajda:90SPA, Chen:96IT, Luschgy&Rukhin&Vajda:93SPA}. Using the
asymptotic distribution of the eigenvalues
 of a Toeplitz matrix\cite{Grenander&Szego:book,Gray:72IT},   we
 have
\begin{equation}
\lim_{n\rightarrow\infty} \frac{1}{n} \log (\det (\Sigmabf_{j,n}))
= \frac{1}{2\pi}\int_0^{2\pi} \log S_{y}^{(j)}(\omega) d\omega,
~~~j=0,1,  \label{eq:GaussianKLvector2}
\end{equation}
where $S_y^{(j)}(\omega)$ is the spectrum of $\{y_i\}$  which is
assumed to have finite lower and upper bounds under distribution
$\ybf_n \sim p_{j,n}$. The limiting behavior of
$n^{-1}\ybf_n^T\Sigmabf_{j,n}^{-1}\ybf_n$ is also known and is
given by (assuming that the true distribution of $\{y_i\}$ is
$p_{0,n}$)
\begin{eqnarray}
\lim_{n\rightarrow\infty} \frac{1}{n}  \ybf_n^T\Sigmabf_{1,n}^{-1}
\ybf_n &=& \frac{1}{2\pi} \int_0^{2\pi}
\frac{S_y^{(0)}(\omega)}{S_y^{(1)}(\omega)}
d\omega, \label{eq:GaussianKLvector31}\\
\lim_{n\rightarrow\infty} \frac{1}{n}  \ybf_n^T\Sigmabf_{0,n}^{-1}
\ybf_n &=& 1, \label{eq:GaussianKLvector3}
\end{eqnarray}
where the limit is in the  almost-sure sense convergence under
$H_0$, provided that $S_y^{(0)}(\omega)$ and $S_y^{(1)}(\omega)$
are continuous and strictly positive. (See Lemma 1 and 2 in
\cite{Hannan:73JAP} and Prop. 10.8.2 and 10.8.3 in
\cite{Brockwell&Davis:book}.) Combining
(\ref{eq:GaussianKLvector})-(\ref{eq:GaussianKLvector3}), the
error exponent for two zero-mean stationary Gaussian processes is
thus given by
\begin{eqnarray}
K &=& \lim_{n\rightarrow \infty} \frac{1}{n}\log
\frac{p_{0,n}}{p_{1,n}}(\ybf_n) ~~~\mbox{a.s.}~[p_{0,n}],\\
 &=&
\frac{1}{2\pi} \int_0^{2\pi} \left(\frac{1}{2} \log
\frac{S_y^{(1)}(\omega)}{S_y^{(0)}(\omega)} +
\frac{S_y^{(0)}(\omega)}{2S_y^{(1)}(\omega)} -\frac{1}{2}\right)
d\omega,\\
&=& \frac{1}{2\pi}\int_0^{2\pi}
 D(\Nc(0,S_y^{(0)}(\omega))||\Nc(0,S_y^{(1)}(\omega)))d\omega.\label{eq:errorexponentspectralnewderiv}
\end{eqnarray}
Intuitively,  the error exponent
(\ref{eq:errorexponentspectralnewderiv})  can be explained from
(\ref{eq:GaussianKLscalar}) using the frequency binning argument
used to obtain the channel capacity of Gaussian channel with
colored noise from that of independent parallel Gaussian channels
\cite{Cover&Thomas:book}.

The spectral form (\ref{eq:errorexponentspectralnewderiv}) of the
error exponent is valid for  a wide class of stationary Gaussian
processes including the autoregressive moving average (ARMA)
processes and the hidden Markov model
(\ref{eq:hypothesisscalarstate}) - (\ref{eq:statespacescalar}).
For the detection (\ref{eq:hypothesisscalarstate}) under the
scalar state space model (\ref{eq:statespacescalar}), we have
\begin{equation} \label{eq:twocovariancemtxs}
\Sigmabf_{0,n} = \Ibf, ~~~~\Sigmabf_{1,n} = \Sigmabf_{s,n} + \Ibf,
\end{equation}
where $\Sigmabf_{s,n} (l,m) = \Pi_0 a^{|l-m|}, ~l,m=1,\cdots,n$,
and two spectra under $H_0$ and $H_1$ are given by
(\ref{eq:spectrumunderH1}). However, it is not straightforward to
obtain a closed-form expression for
(\ref{eq:errorexponentspectralnewderiv}) except in some special
cases, e.g., when both of the two distributions of $\{y_i\}$ under
$H_0$ and $H_1$ have the Markov property \cite{Luschgy:94SJS}.

In the remainder of the paper, we focus on the derivation of  a
closed-form expression for the error exponent
 $K$
 of the miss probability  for (\ref{eq:hypothesisscalarstate}) - (\ref{eq:statespacescalar}) by exploiting the
state-space structure under the alternative hypothesis. We do so
by making a connection with Kalman filtering\cite{Kailath:70PROC}.
Our expressions will allow us to investigate the properties of the
error exponent.

\subsection{Closed-Form Error Exponent via Innovations Approach}\label{subsec:errorexponentinnovation}

\vspace{0.5em}
\begin{theorem}[Error exponent] \label{theo:errorexponentNPscalarstate}
For the Neyman-Pearson detector for the hypotheses
(\ref{eq:hypothesisscalarstate}) with level $\alpha \in (0,1)$
(i.e. $ P_F \le \alpha$) and $0 \le a \le 1$, the  error exponent
of the miss probability is given by
\begin{eqnarray}
K &=& \frac{1}{2} \log\frac{R_e}{\sigma^2} + \frac{1}{2}
\frac{\tilde{R}_e}{R_e}
 - \frac{1}{2}, \label{eq:errorexponentscalar}
\end{eqnarray}
independently of the value of $\alpha$, where $R_e$
 and $\tilde{R}_e$ are the steady-state
  variances of the innovations process of $\{y_i\}$ calculated under $H_1$ and $H_0$, respectively.
  Specifically, $R_e$ and $\tilde{R}_e$ are given by
\begin{equation}
R_e = P + \sigma^2, \label{eq:Reinfexplicit}
\end{equation}
and
\begin{equation}
\tilde{R}_e
=\sigma^2\left(1+\frac{a^2P^2}{P^2+2\sigma^2P+(1-a^2)\sigma^4}\right),
\label{eq:tReinfexplicit}
\end{equation}
where
\begin{equation}
 P =
\frac{1}{2}\sqrt{[\sigma^2(1-a^2)-Q]^2+4\sigma^2
Q}-\frac{1}{2}\sigma^2(1-a^2) + \frac{Q}{2}. \label{eq:Pinfexplicit}
\end{equation}
Here, $P$ is the steady-state error variance of the minimum mean
square error (MMSE) estimator  for the signal $s_i$ under the
model $H_1$.
   Note that the error exponent
(\ref{eq:errorexponentscalar}) is thus a closed-form of
(\ref{eq:errorexponentspectralnewderiv}) for the state-space
model.
\end{theorem}

 \vspace{0.5em}
{\em Proof:} See the Appendix. \vspace{0.5em}

Theorem \ref{theo:errorexponentNPscalarstate} follows from the
fact that the almost-sure limit (\ref{eq:asymKLrate}) of the
normalized log-likelihood ratio under $H_0$
 is the  error exponent for general ergodic
 cases\cite{Vajda:book, Vajda:90SPA, Chen:96IT,  Luschgy&Rukhin&Vajda:93SPA}.
 To make the closed-form calculation of the error exponent tractable
 for the hidden Markov structure of $\{y_i\}$,
 we express the log-likelihood ratio through
 the innovations representation \cite{Schweppe:65IT};
 the log-likelihood ratio is given by a function of the
 sum of squares of the innovations on which  the strong
 law of large numbers (SLLN) is  applied.  The calculated innovations are true in the sense that
 they form an independent sequence only under $H_1$, i.e.,
 when the signal actually comes from the state-space model.
It is worth noting that  $\tilde{R}_e$ is the steady-state
variance of the ``innovations'' calculated as if the observations
result
   from the alternative, but are actually from the null hypothesis.
   In this case, the ``innovation'' sequence becomes the output of
    a recursive (whitening) filter driven by an i.i.d. process $\{y_i\}$ since
    the Kalman filter converges to the recursive Wiener filter
     for time-invariant stable systems \cite{Kailath&Sayed&Hassibi:book}.

The relationship between the spectral-domain approach and the
innovations approach is  explained by the canonical spectral
factorization,  which is well established for the state-space
model.  The asymptotic variance of the innovations sequence is the
key parameter in both cases.  The relationship between the
asymptotic performance of the Neyman-Pearson detector and that of
the Kalman filter  is evident in (\ref{eq:errorexponentscalar})
for the state-space model. In both cases, the innovations process
plays a critical role in characterizing the performance, and the
asymptotic
      variance of the innovation is sufficient for
       the calculation of the error exponent for the Neyman-Pearson detector and the steady-state
       error variance for the Kalman filter.

\section{Properties of Error Exponent}\label{sec:errorexp_properties}

In this section, we  investigate  the properties of the error
exponent  derived in the preceding section.  We particularly examine
the large sample error behavior
 with respect to the correlation strength and SNR.  We show that the intensity of the additive
 noise significantly changes the error behavior with respect to the correlation strength,
 and the error exponent has a distinct phase transition in behavior with respect to the
 correlation strength depending on SNR.

\vspace{0.5em}
\begin{theorem}[$K$ vs. correlation]  \label{theo:etavscorrelation}
The error exponent $K$ is a continuous function of the correlation
coefficient $a$ ($0 \le a \le 1$) for a given
 SNR $\ge 0$. The error exponent as a function of correlation strength is
characterized by the following:
\begin{itemize}
\item[(i)]  For SNR  $\Gamma \ge 1$, $K$ is monotonically
decreasing as the correlation strength increases (i.e. $a \uparrow
1$); \item[(ii)] For SNR $\Gamma < 1$, there exists a non-zero
value $a^*$ of the correlation coefficient  that achieves the
maximal $K$, and $a^*$ is given by the solution of the following
equation.
\begin{equation} \label{eq:stationarypoint}
[1+a^2+\Gamma(1-a^2)]^2-2\left(r_e+\frac{a^4}{r_e}\right)=0,
\end{equation} where $r_e = R_e/\sigma^2$. Furthermore, $a^*$ converges to one
as $\Gamma$ goes to zero.
\end{itemize}
\end{theorem}

\vspace{0.5em} {\em Proof:} See the Appendix. \vspace{0.5em}

\begin{figure}[htbp]
\centerline{
{
    \begin{psfrags}
    \scalefig{0.55}\epsfbox{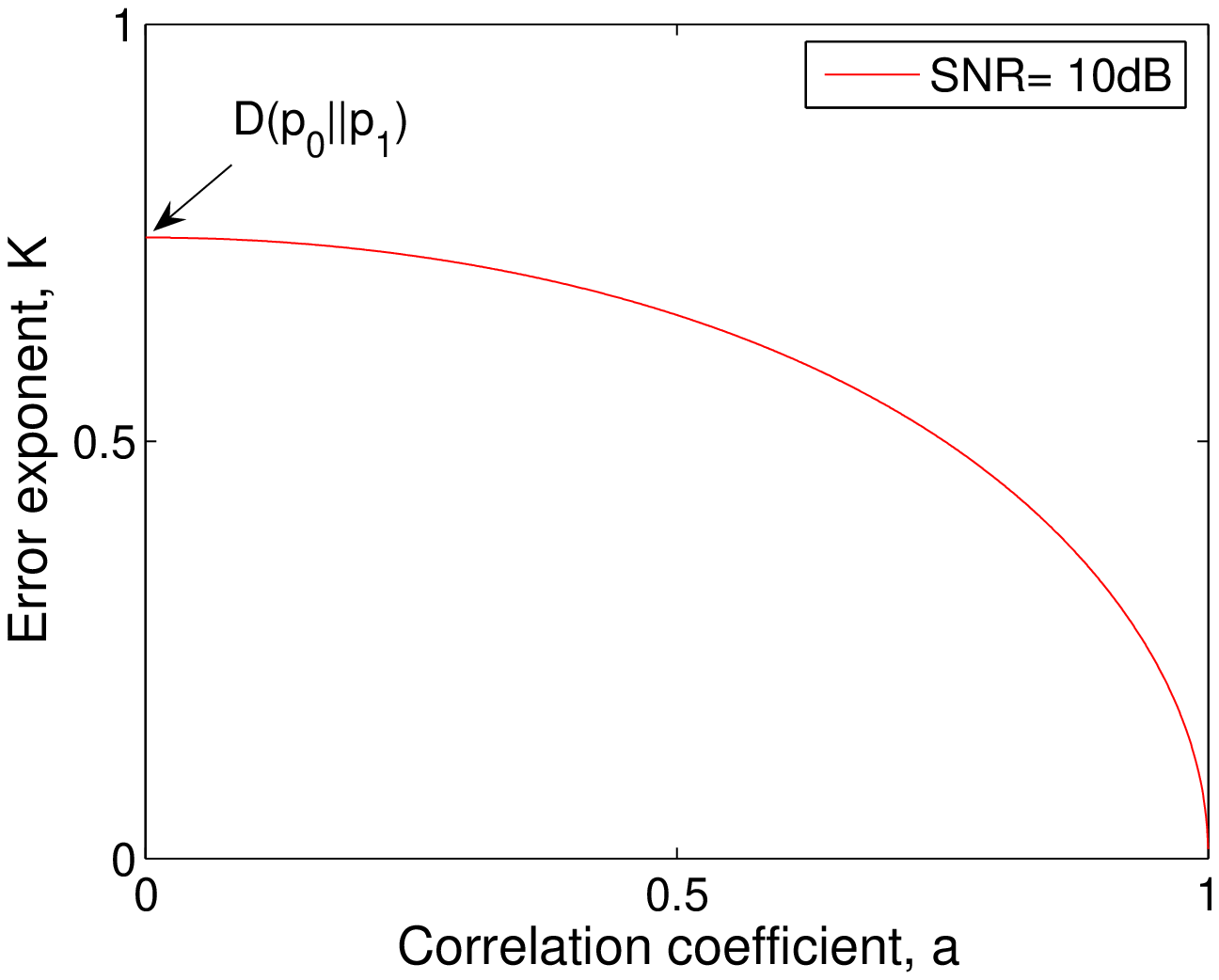}
    \end{psfrags}
} } \caption{$K$ versus correlation coefficient $a$ (SNR=10 dB):
$p_0 = \Nc(0,\sigma^2), ~p_1=\Nc(0,\Pi_0+\sigma^2)$}
\label{fig:snr10dBetavsa}
\end{figure}

We first note that  Theorem \ref{theo:etavscorrelation} shows that
an  i.i.d. signal gives the best error performance for a given SNR
$\ge 1$ with the maximal  error exponent being
$D(\Nc(0,\sigma^2)||\Nc(0,\Pi_0+\sigma^2))$. (In this case,
Theorem \ref{theo:errorexponentNPscalarstate} reduces to  Stein's
lemma.)
 The intuition behind this result is that  the signal component in the observation is strong at high SNR,
 and the innovations (the new information) provide more benefit to the detector than the noise averaging effect
 present for correlated observations. That is, simple radiometry
 provides sufficient detection power when the signal level is
 above that of the noise.
 Fig. \ref{fig:snr10dBetavsa} shows the error exponent as a function of the correlation
 coefficient $a$ for SNR $\Gamma =$ 10 dB.
 The monotonicity of the error exponent is clearly seen; moreover, we see that
 the amount of decrease becomes larger as $a$ increases.
 Notice also that the amount of performance degradation
 from the i.i.d. case is not severe for weak correlation and the error exponent decreases suddenly near $a=1$
and eventually becomes zero
 at $a=1$. (It is easy to show that the miss probability
 decays with $\Theta(\frac{1}{\sqrt{n}})$ for any SNR at $a=1$.)

\begin{figure}[htbp]
\centerline{
{
    \begin{psfrags}
    \scalefig{0.55}\epsfbox{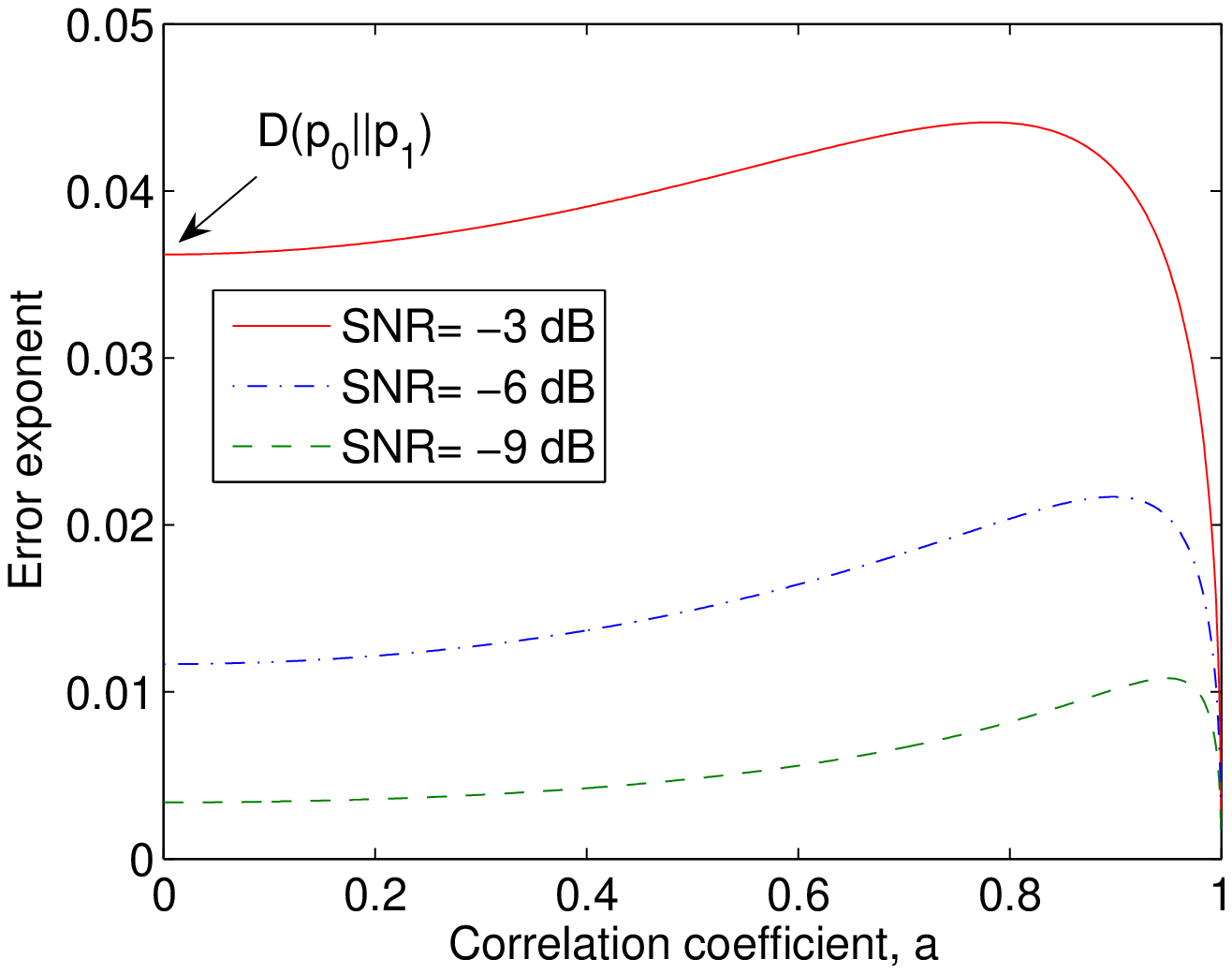}
    \end{psfrags}
}
}
\caption{$K$ versus correlation coefficient $a$ (SNR= -3, -6, -9 dB)}
\label{fig:snrm3m6dBetavsa}
\end{figure}

 In contrast, the error exponent  does not decrease  monotonically in $a$ for SNR $< 1$,
 and there exists an optimal  correlation as shown in Fig. \ref{fig:snrm3m6dBetavsa}.
 It is seen that the i.i.d. case no longer gives the  error performance for a fixed SNR.
 The error exponent initially increases as $a$ increases, and then decreases to zero as $a$ approaches one.
 As the SNR further decreases (see the cases of -6 dB and -9dB) the error exponent decreases for a fixed
  correlation strength, and  the value of $a$ achieving the maximal error exponent is shifted closer to one.
At low SNR the noise in the observation dominates.  So,
intuitively, making the signal more correlated provides greater
benefit of noise averaging. The lower the SNR, the stronger we
would like the correlation to be in order to compensate for the
dominant noise power, as shown in Fig.~\ref{fig:snrm3m6dBetavsa}.
However, excessive correlation in the signal does not provide new
information by observation, and the error exponent ultimately
converges to zero as $a$ approaches one. Notice that the ratio of
the error exponent for the optimal correlation to that for the
i.i.d. case becomes large as SNR decreases. Hence, the improvement
due to optimal correlation can be large for low SNR cases.
\begin{figure}[htbp]
\centerline{
{
    \begin{psfrags}
    \scalefig{0.55}\epsfbox{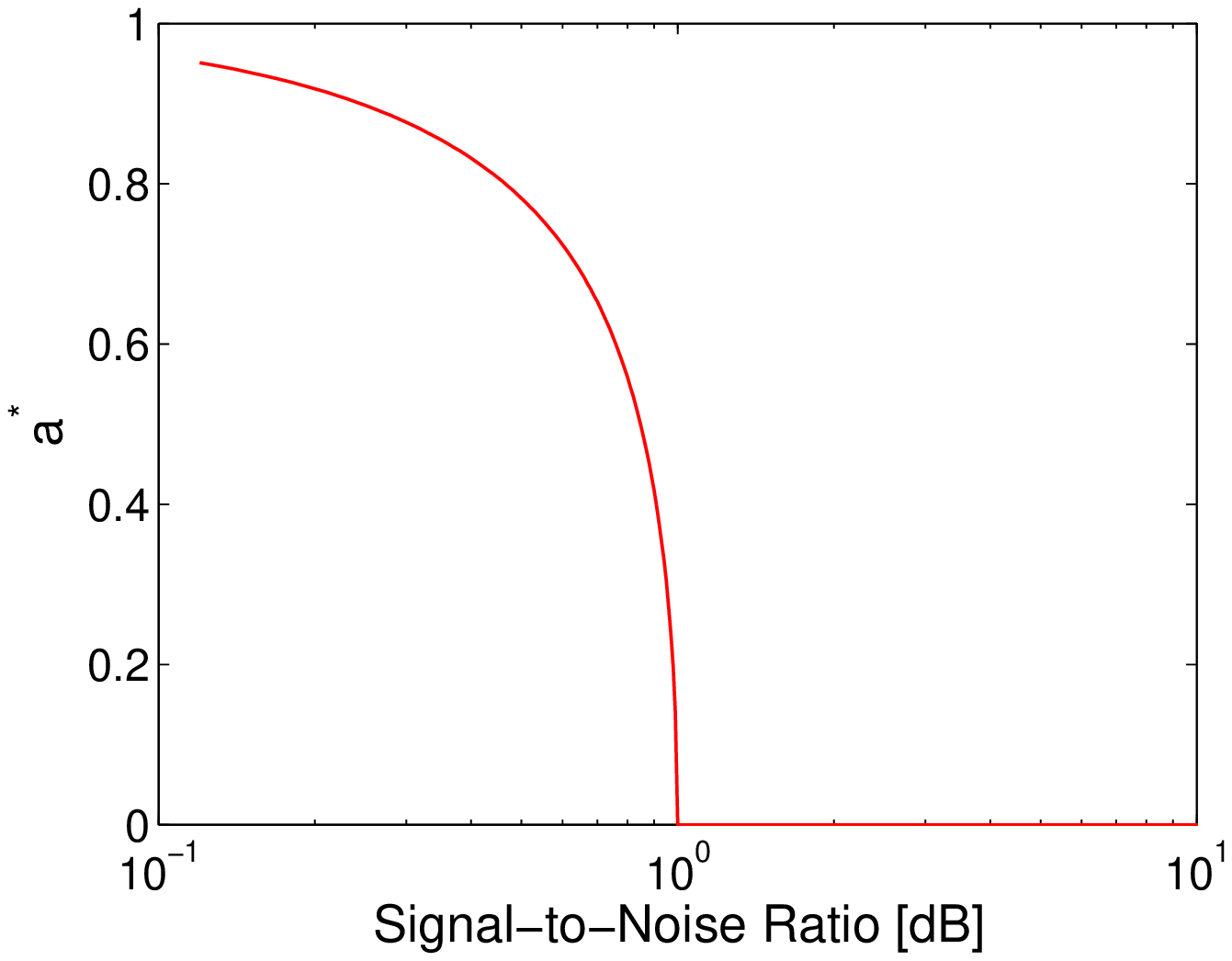}
    \end{psfrags}
}
}
\caption{Optimum correlation strength  versus SNR}
\label{fig:amaxvssnr}
\end{figure}
  Fig. \ref{fig:amaxvssnr} shows the value of  $a$ that maximizes
  the error exponent as a function of SNR.  As shown in the figure,
   unit SNR is a transition point between two different behavioral regimes  of
   the error  exponent with respect to correlation strength, and the transition is
    very sharp; the optimal correlation strength $a$ approaches one rapidly once
SNR becomes smaller than one.

The behavior of the error exponent with respect to SNR is given by
the following theorem.

\vspace{0.5em}
\begin{theorem}[$K$  vs. SNR] \label{theo:etavsSNR}
The error exponent  $K$ is monotonically  increasing as SNR
increases for a given correlation coefficient $0 \le a <1$.
Moreover, at high SNR the error exponent $K$ increases linearly
with respect to $\frac{1}{2}\log [1+\mbox{SNR}(1-a^2)]$.
\end{theorem}

\vspace{0.5em} {\em Proof:} See the Appendix. \vspace{0.5em}

\begin{figure}[htbp]
\centerline{
{
    \begin{psfrags}
    \scalefig{0.55}\epsfbox{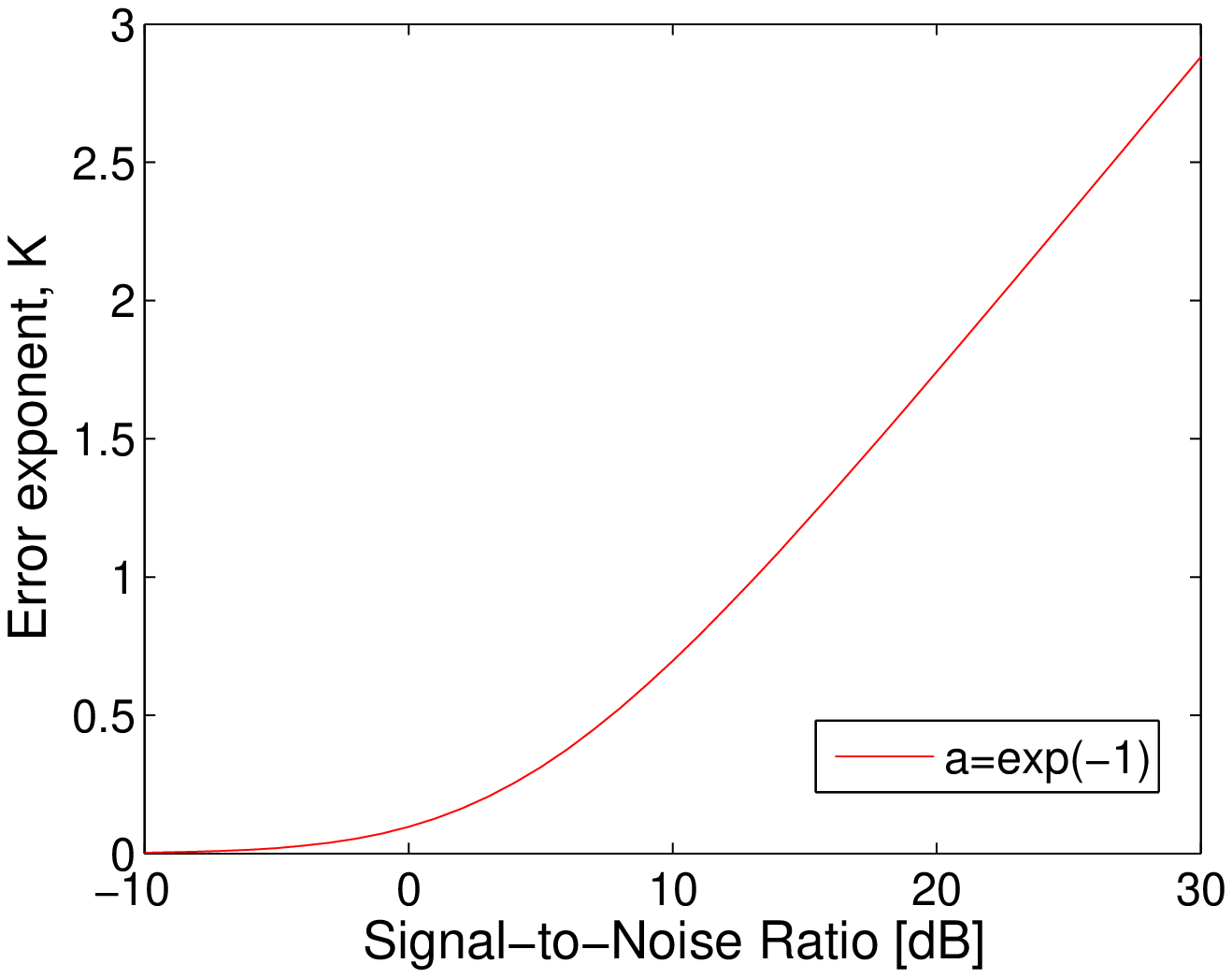}
    \end{psfrags}
}
}
\caption{$K$ versus SNR ($a=e^{-1}$)}
\label{fig:etavsSNR1}
\end{figure}

  The
detrimental effect of correlation at high SNR is
clear.\footnote{Interestingly, the error exponent at high SNR has
the same expression as the capacity of the Gaussian channel.} The
performance degradation due to correlation is equivalent to the
SNR decreasing by factor $(1-a^2)$. The $\log (1+SNR)$ increase of
$K$ w.r.t. SNR is analogous to similar error-rate behavior arising
in diversity combining of versions of a communications signal
arriving over independent Rayleigh-faded
 paths in additive noise, where the error probability is given by $
P_e \sim (1+\mbox{SNR})^{-L}$ and $L$ is the number of independent
multipaths.  In both cases, the signal component is random. The
$\log$ SNR behavior of the optimal Neyman-Pearson detector for
stochastic signals applies to general correlations as well with a
modified definition of SNR. Comparing with the detection of a
deterministic signal in noise, where the error exponent is
proportional to SNR, the increase of error exponent w.r.t. SNR is
much slower for the case of a stochastic signal in noise.  Fig. 4
shows the error exponent with respect to SNR for a given
correlation strength. The $\log$  SNR behavior is evident at high
SNR.

\section{Extension to The Vector Case} \label{sec:vectormodel}

In order to treat general cases in which the signal is a higher
order AR process or the signal is determined by a linear combination
of several underlying phenomena, we now consider a vector
state-space model, and extend the results of the previous sections
to this model. The hypotheses for the vector case are given by
\begin{equation}  \label{eq:hypothesisvector}
\begin{array}{lcl}
H_0 &: & y_i = w_i, ~~~~~~~~~~~~~~~i=1,2,\cdots, n, \\
H_1 &: & y_i = \hbf^T \sbf_i+ w_i,\\
\end{array}
\end{equation}
where  $\hbf$ is a known vector and $\sbf_i \defeq
[s_{1i},s_{2i},\cdots, s_{mi}]^T$ is the state of an
$m$-dimensional process at time $i$ following the state-space
model
\begin{eqnarray} \label{eq:statespacevector}
\sbf_{i+1}&=&\Abf \sbf_i + \Bbf \ubf_i, \\
\sbf_1 &\sim& \Nc({\mathbf 0},~{\mathbf \Pi}_0), \nn \\
\ubf_i &\stackrel{i.i.d.}{\sim}& \Nc({\mathbf 0}, ~\Qbf), ~~\Qbf \ge 0. \nn
\end{eqnarray}
We assume that the feedback and input matrices,  $\Abf$ and
$\Bbf$, are known with $|\lambda_k(\Abf)| < 1$ for all $k$, and
the process noise $\{\ubf_i\}$ independent of the measurement
noise $\{w_i\}$. We also assume that the initial state $\sbf_1$ is
independent of $\ubf_i$ for all $i$, and the initial covariance
$\Pibf_0$ satisfies the following Lyapunov equation
\begin{equation} \Pibf_0 = \Abf \Pibf_0 \Abf^T + \Bbf\Qbf\Bbf^T.
\end{equation} Thus, the signal sequence $\{\sbf_i\}$ forms a
stationary vector process. In this case the SNR is defined
similarly to (\ref{eq:SNRDEFscalar}) as
$\frac{\hbf^T\Pibf_0\hbf}{\sigma^2}$. The autocovariance of the
observation sequence $\{y_i\}$ is given by
\begin{equation}
r_y(i,j)= \Ebb \{y_i y_j \} = \left\{
\begin{array}{ll}
\sigma^2 \delta_{ij}   & \mbox{under}~ H_0, \\
\hbf^T \Abf^{|i-j|}\Pibf_0 \hbf + \sigma^2\delta_{ij}   & \mbox{under}~ H_1,
\end{array}
\right.
\end{equation}
where $\delta_{ij}$ is the Kronecker delta. Thus, the covariance
matrix of the observation under $H_1$ is a symmetric Toeplitz
matrix with $\hbf^T \Abf^l \Pibf_0 \hbf$ as the $l$th off-diagonal
entry ($l \ne 1$).  Since $|\lambda_k(\Abf)| < 1$ for all $k$,
$c_l
\defeq \hbf^T\Abf^l \Pibf_0 \hbf$ is an absolutely summable
sequence and the eigenvalues of the covariance matrix of $\ybf_n$
is bounded both from below and from above.

\vspace{0.5em}
\begin{theorem}[Error exponent] \label{theo:errorexponentNPvector}
For the Neyman-Pearson detector for the hypotheses
(\ref{eq:hypothesisvector}) and (\ref{eq:statespacevector}) with
level $\alpha \in (0,1)$ (i.e. $ P_F \le \alpha$) and
$|\lambda_k(\Abf)| < 1$ for all $k$, the  error exponent of the
miss probability is given by (\ref{eq:errorexponentscalar})
independently of the value of $\alpha$. The steady-state
  variances of the innovation process $R_e$
 and $\tilde{R}_e$ calculated under $H_1$ and $H_0$, respectively,
 are given by
\begin{equation}
R_e =\sigma^2 + \hbf^T \Pbf \hbf,
\end{equation}
where $\Pbf$ is the  unique stabilizing solution of  the
discrete-time algebraic Riccati equation
\begin{equation}  \label{eq:RiccatiVector}
\Pbf = \Abf \Pbf \Abf^T + \Bbf\Qbf\Bbf^T - \frac{\Abf \Pbf \hbf \hbf^T \Pbf \Abf^T}{ \hbf^T \Pbf \hbf+\sigma^2},
\end{equation}
and
\begin{equation}
\tilde{R}_e = \sigma^2(1+\hbf^T \tilde{\Pbf} \hbf),
\end{equation}
where $\tilde{\Pbf}$ is the unique positive-semidefinite solution of
the following Lyapunov equation
\begin{equation}  \label{eq:theo1Lyapunov}
\tilde{\Pbf}= (\Abf - \Kbf_p \hbf^T) \tilde{\Pbf}(\Abf-\Kbf_p \hbf^T)^T + \Kbf_p \Kbf_p^T,
\end{equation}
and $\Kbf_p =  \Abf\Pbf\hbf R_e^{-1}$.

In spectral form,  $K$ is given by
(\ref{eq:errorexponentspectralnewderiv}), where
$S_y^{(0)}(\omega)=\sigma^2 ~(-\pi \le \omega \le \pi)$ and
$S_y^{(1)}(\omega)$ is given by
\begin{equation} \label{eq:spectraldensityvector}
S_y^{(1)}(\omega)=[ \hbf^T(e^{j\omega}\Ibf - \Abf)^{-1} ~~1]
\left[
\begin{array}{cc}
\Qbf & 0 \\
0 & \sigma^2
\end{array}
\right]
\left[
\begin{array}{c}
(e^{-j\omega}\Ibf -\Abf^T)^{-1}\hbf \\
1
\end{array}
\right].
\end{equation}
\end{theorem}

\vspace{0.5em} {\em Proof:} See the Appendix. \vspace{0.5em}

 For this vector model, simple results describing the properties of the error exponent
  are not tractable since the relevant expressions  depend on the multiple eigenvalues of the
  matrix $\Abf$.
  However, (\ref{eq:errorexponentscalar}), (\ref{eq:RiccatiVector}) and (\ref{eq:theo1Lyapunov}) provide
   closed-form expressions for the error exponent which can easily be
    explored numerically.

\section{Simulation Results}
\label{sec:numerical}

To verify the behavior of the miss probability predicted by our
asymptotic analysis, in this section we provide some simulation
results. We consider the scalar model (\ref{eq:statespacescalar}),
for SNR of 10 dB and - 3 dB, and for several correlation
strengths. The probability of false alarm is  set at 0.1\% for all
cases we consider.
\begin{figure}[htbp]
\centerline{
    \begin{psfrags}
    \scalefig{0.55}
    \epsfbox{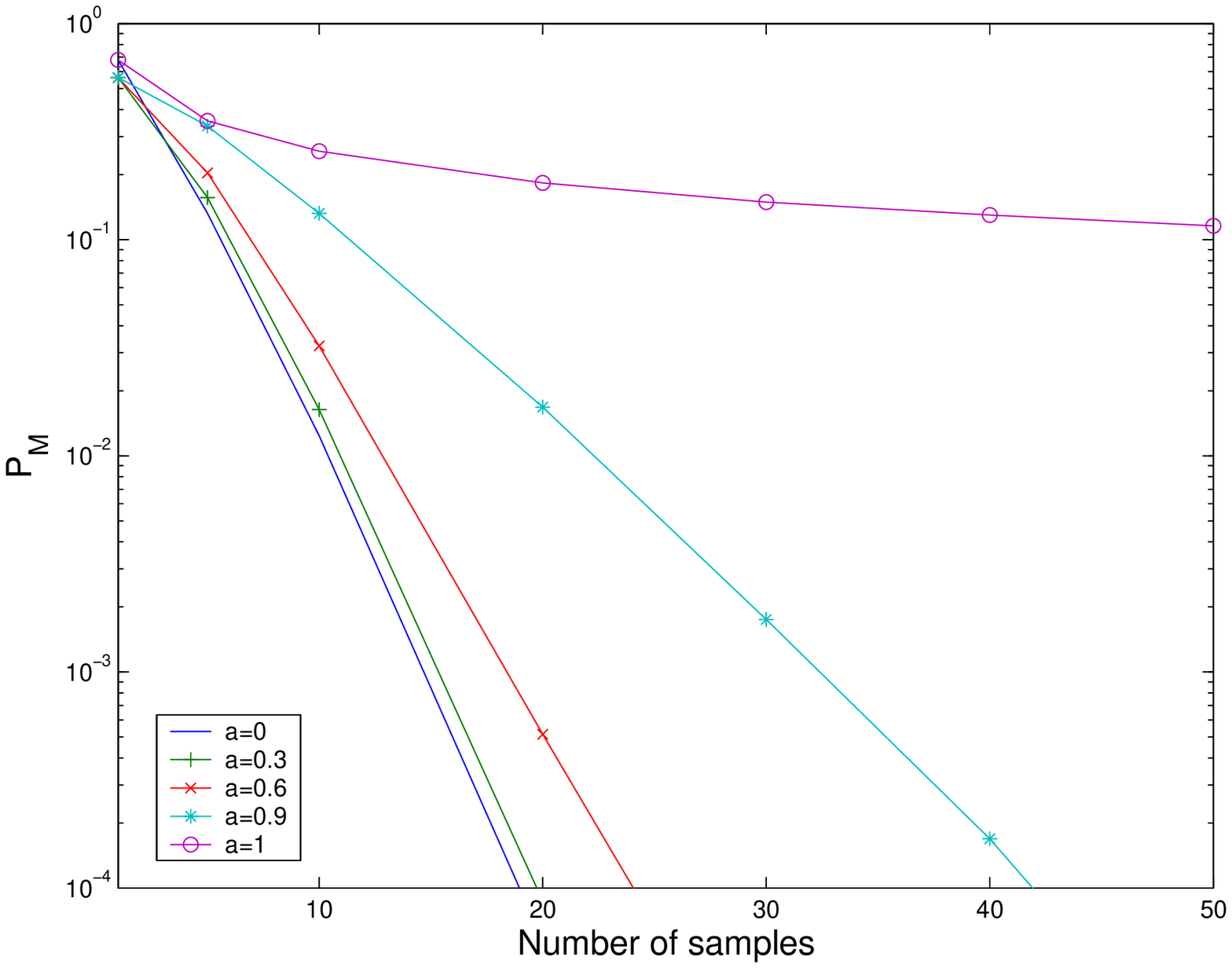}
    \end{psfrags}
}
\caption{ $P_M$ vs.  number of samples (SNR=10dB)}
\label{fig:PMvsnosensors10dB}
\end{figure}

Fig. \ref{fig:PMvsnosensors10dB} shows the simulated miss
probability as a function of the number of samples for 10 dB SNR.
It is seen, as predicted by our analysis, that the i.i.d. case
($a=0$) has the largest slope for error decay, and the slope is
monotonically decreasing as $a$ increases to one. Notice that the
error performance for the same number of observations is
significantly different for different correlation strengths for
the same SNR, and the performance for weak correlation is not much
different from the i.i.d. case, as predicted by Fig.
\ref{fig:snr10dBetavsa}.  It is also seen that the miss
probability for the perfectly correlated case ($a=1$) is not
exponentially decaying, again confirming our analysis.

\begin{figure}[htbp]
\centerline{
    \begin{psfrags}
    \scalefig{0.55}
    \epsfbox{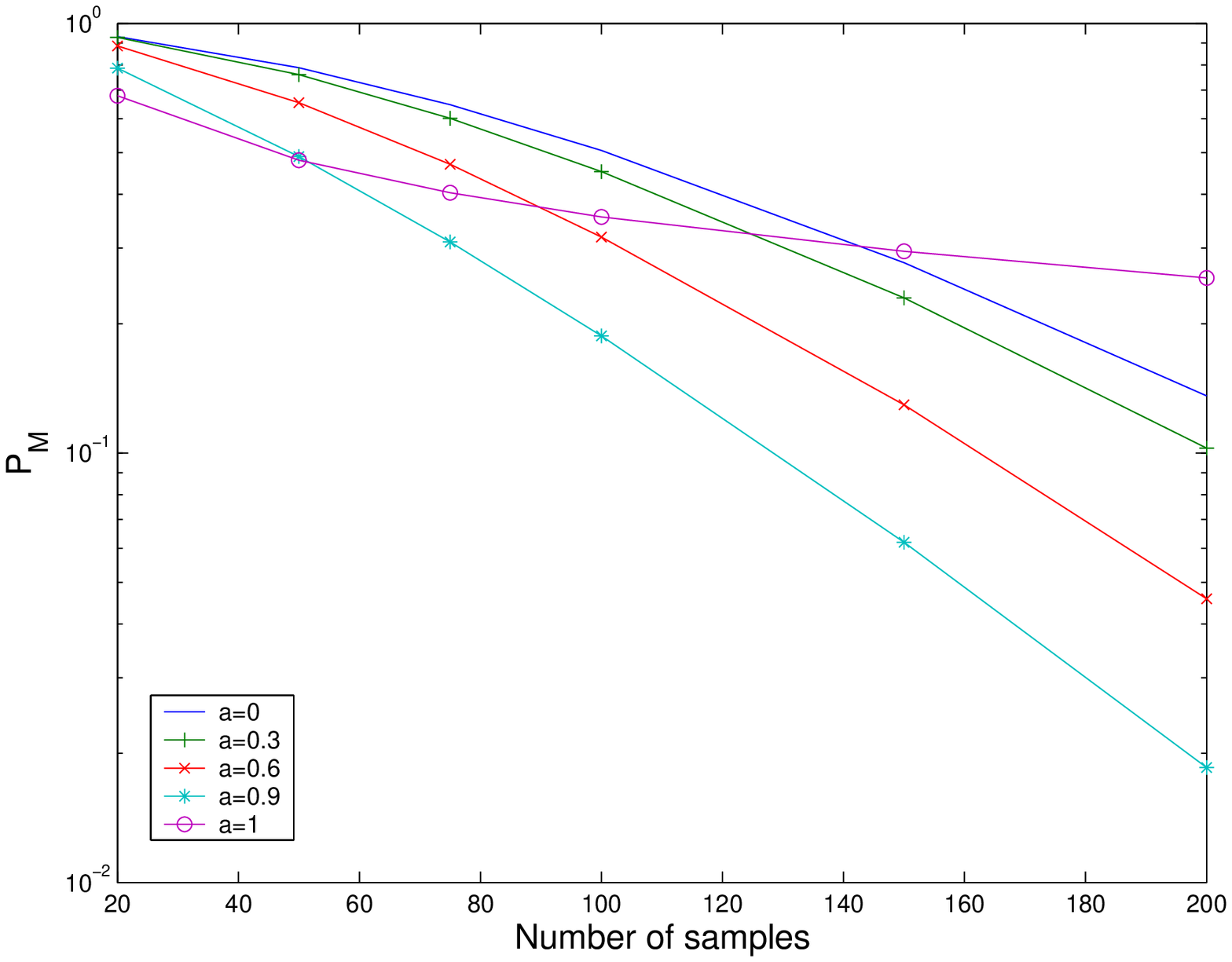}
    \end{psfrags}
}
\caption{ $P_M$ vs.  number of samples (SNR=-3dB)}
\label{fig:PMvsnosensorsm3dB}
\end{figure}

The simulated error performance for SNR of -3 dB is shown in Fig.
\ref{fig:PMvsnosensorsm3dB}. It is seen that the asymptotic slope
of $\log P_M$ increases as $a$ increases from zero as predicted by
Theorem \ref{theo:etavscorrelation}, and reaches a maximum  with a
sudden decrease after the maximum. Notice that the error curve is
still not a straight line for the low SNR case due to the $o(n)$
term in the exponent of the error probability.  Since the error
exponent increases only with $\log$ SNR, the required number of
observations for -3 dB SNR  is much larger than for 10 dB SNR for
the same miss probability. It is clearly seen that $P_M$ is still
larger than $10^{-2}$  for 200 samples whereas it is $10^{-4}$
with 20 samples for the 10 dB SNR case.

\section{Conclusion}
\label{sec:conclusion}

We have considered the detection of correlated random signals
using  noisy observations.  We have derived the  error exponent
for
 the Neyman-Pearson detector of a fixed level using the  spectral domain and the innovations approaches.
   We have also provided the error exponent in closed form for the vector state-space model.
The closed-form expression is valid not only for the state-space
model but also for any orthogonal transformation of the original
observations under the state-space model, since the spectral
domain result does not change by orthogonal transformation and
 Theorem 1 is a
closed-form expression of the invariant spectral form. We have
investigated the properties of the error exponent for the scalar
case. The error exponent is a function of SNR and correlation
strength. The behavior of the error exponent as a function of
correlation strength is sharply divided into two regimes depending
on SNR. For SNR $\ge~ 1$ the error exponent is monotonically
decreasing in the signal correlation. On the other hand, for SNR
$<~ 1$, there is a non-zero correlation strength that gives the
maximal error exponent. Simulations confirm the validity of our
asymptotic results for finite sample sizes.

\section*{Acknowledgement}
The authors wish to thank an anonymous reviewer for pointing out a
simple derivation of the spectral domain result in Section
\ref{sec:errorexponentCF}.

\section*{Appendix}

{\em Proof of Theorem~\ref{theo:errorexponentNPscalarstate}}

Since the error exponent for the Neyman-Pearson detector with a
fixed level $\alpha \in (0,1)$ is given by the almost-sure limit
of the normalized log-likelihood ratio $\frac{1}{n}\log
L_n(\ybf_n)$ under $H_0$ (if the limit
exists)\cite{Vajda:book,Vajda:90SPA,Luschgy&Rukhin&Vajda:93SPA,Chen:96IT},
we focus on the calculation of the limit.
 We show that $\frac{1}{n}\log L_n$ converges a.s. under $H_0$ for Gauss-Markov  signals in noise
 using the limit distribution of the innovations sequence. The log-likelihood ratio is given by
\begin{equation} \label{eq:ProofKInnovloglike}
\log L_n(\ybf_n) = \log p_{1,n}(\ybf_n)-\log p_{0,n}(\ybf_n).
\end{equation}
 We have, for the second term on the right-handed side (RHS) of
(\ref{eq:ProofKInnovloglike}),
\[
p_{0,n} (\ybf_n) =\frac{1}{ (2\pi\sigma^2)^{n/2}}e^{-\frac{1}{2}\sum_{i=1}^{n} y_i^2/\sigma^2},
\]
and so, under $H_0$,
\begin{eqnarray}
\frac{1}{n}\log p_{0,n}(\ybf) &=&  -\frac{1}{2}\log (2\pi\sigma^2)
-\frac{1}{2\sigma^2 n}\sum_{i=1}^{n} y_i^2, \label{eq:ProofKInnovp0}\\
&\rightarrow&  -\frac{1}{2}\log (2\pi\sigma^2) -\frac{1}{2}
~~~\mbox{a.s.},
\end{eqnarray}
since   $\frac{1}{n}\sum_{i=1}^n y_i^2 \rightarrow \Ebb_0
\{Y_1^2\} = \sigma^2$ a.s. under $H_0$ as $n\rightarrow\infty$ by
the SLLN. Now consider the first term on the RHS of
(\ref{eq:ProofKInnovloglike}).
 The log-likelihood  under $H_1$  can be obtained  via the  Kalman recursion for the
 innovations \cite{Helstrom:book},\cite{Schweppe:65IT}.  Specifically, define $l_i \defeq \log p_{1,i} (y_1,y_2,\cdots,y_i)$ and
 $y_1^i \defeq \{y_1,y_2,\cdots, y_i\}$. Then,
\begin{equation}
 p_{1,i}(y_1^i)=p_{1,i}(y_1^{i-1})p_{1,i}(y_i|y_1^{i-1}), ~~~2\le i
 \le n.
 \end{equation}
  Hence,
  \begin{equation}
  l_i = l_{i-1} + \log p_{1,i}(y_i|y_1^{i-1}), ~~~2\le i
 \le n,
  \end{equation}
  where $l_1 = \log p_{1,1}(y_1)$.
   Since the joint
distribution of $\{y_1,y_2,\cdots,y_i\}$ is Gaussian,  the
conditional distribution $p_{1,i}(y_i|y_1^{i-1})$ is also Gaussian
with mean $\hat{y}_{i|i-1}$ and variance $R_{e,i}$. $l_i$ is
expressed using the innovations representation by
\begin{equation} l_i =
l_{i-1} -\frac{1}{2}\log (2\pi R_{e,i}) - \frac{1}{2}
\frac{e_i^2}{R_{e,i}},
\end{equation}
where the minimum mean square error (MMSE) prediction
$\hat{y}_{i|i-1}$  of $y_i$ is the conditional expectation $\Ebb_1
\{y_i|y_1^{i-1}\}$ and the innovation is given by $e_i \defeq y_i
- \hat{y}_{i|i-1}$ with  variance $R_{e,i} = \Ebb_1 \{e_i^2\}$.
Hence,
\begin{equation}\label{eq:ProofKInnovfirstterm}
\frac{1}{n}\log p_{1,n}(\ybf_n) = \frac{l_n}{n} = -\frac{1}{2}\log
(2\pi) - \frac{1}{2n}\sum_{i=1}^{n}\log R_{e,i} -
\frac{1}{2n}\sum_{i=1}^{n} \frac{e_i^2}{R_{e,i}}.
\end{equation}
The second term on the RHS of (\ref{eq:ProofKInnovfirstterm}) is not
random, and we have
\begin{equation}
\frac{1}{2n}\sum_{i=1}^{n}\log R_{e,i}  \rightarrow \frac{1}{2}\log
R_e,
\end{equation}
by the  Ces\'{a}ro mean theorem since $R_{e,i} \rightarrow R_e$
and $R_{e,i} \ge \sigma^2 >0$ for all $i$ where $R_e$ is given by
\begin{equation} \label{eq:proofKInnovReinfty}
R_e =  P  + \sigma^2,
\end{equation}
and where $P$ is the steady-state error variance of the optimal
one-step predictor for the signal $\{s_i\}$. Now representing $e_i$
as a linear combination of $y_1,y_2,\ldots, y_i$ gives
\begin{eqnarray}
e_i &=& ~y_i -K_{p,i-1}y_{i-1} -(a-K_{p,i-1})K_{p,i-2}y_{i-2} -(a-K_{p,i-1})(a-K_{p,i-2})K_{p,i-3}y_{i-3}\nn\\
    && ~~-(a-K_{p,i-1})(a-K_{p,i-2})(a-K_{p,i-3})K_{p,i-4}y_{i-4} - \cdots,
\end{eqnarray}
where $K_{p,i} \defeq a P_i R_{e,i}^{-1}$ is the Kalman prediction
gain,  $P_i\defeq\Ebb_1 \{(s_i -\hat{s}_{i|i-1})^2\}$ is the error
variance at time $i$, $\hat{s}_{i|i-1}$ is the linear MMSE
prediction of $s_i$ given $y_1^{i-1}$. Since the Kalman filter
converges asymptotically to the time-invariant recursive Wiener
filter for $0\le a < 1$, we have asymptotically
\begin{eqnarray}
e_i &=& ~y_i -K_{p}y_{i-1} -(a-K_{p})K_{p}y_{i-2} -(a-K_{p})
(a-K_{p})K_{p}y_{i-3}\nn\\
    && ~~-(a-K_{p})(a-K_{p})(a-K_{p})K_{p}y_{i-4} - \cdots,
\end{eqnarray}
where $K_{p}$ is the steady-state Kalman prediction gain. Thus,
under $H_0$ the innovations sequence becomes the output of a stable
recursive filter driven by an i.i.d. sequence $\{y_i\}$, and it is
known to be an ergodic sequence. By the ergodic theorem,
$\frac{1}{n}\sum_{i=1}^n e_i^2$ converges to the true expectation,
which is given by
\begin{equation}
\tilde{R}_e=  \lim_{i\rightarrow\infty} \Ebb_0 \{e_i^2\} =
\sigma^2\left(1+\frac{K_{p}^2}{1-(a-K_{p})^2}\right)
\end{equation}
since $\{y_i\}_{i=1}^\infty$ is an independent sequence under $H_0$.
Substituting $K_{p}$ and $R_e$, we have
\begin{equation}
\tilde{R}_e = \sigma^2\left(1+
\frac{a^2P^2}{R_e^2-a^2\sigma^4}\right)=\sigma^2\left(1+\frac{a^2P^2}{P^2+2\sigma^2P+(1-a^2)\sigma^4}\right).
\end{equation}
 Now, the last term on the RHS of (\ref{eq:ProofKInnovfirstterm}) is given by
\begin{eqnarray}
\frac{1}{2n}\sum_{i=1}^{n} \frac{e_i^2}{R_{e,i}}
&=& \frac{1}{2n}\sum_{i=1}^{n}\frac{e_i^2}{R_e} \frac{R_e}{R_{e,i}}=\frac{1}{2n}\sum_{i=1}^{n}\frac{e_i^2}{R_e} \frac{R_e+C\epsilon^n - C\epsilon^n}{R_e+C\epsilon^n},\\
&=&\frac{1}{2nR_e}\sum_{i=1}^{n}e_i^2 -
\frac{1}{2nR_e}\sum_{i=1}^{n} e_i^2
\frac{C\epsilon^n}{R_e+C\epsilon^n},
\label{eq:ProofKInnovlastline}
\end{eqnarray}
where some positive constant $C$ and $|\epsilon| < 1$  by the
exponential convergence of $R_{e,i}$ to $R_e$. The first term in
(\ref{eq:ProofKInnovlastline}) converges to
$\frac{\tilde{R}_e}{2R_e}$ and
 the second term converges to zero since $\sum_{i=1}^n e_i^2/n$ converges
 to a finite constant and $R_e \ge \sigma^2 > 0$.
 Hence,
(\ref{eq:errorexponentscalar}) follows for $0\le a < 1$.  When
$a=1$, we have $P_M \sim \Theta\left( \frac{1}{\sqrt{n}} \right)$
and $K=0$. We also have $P=0$, $\tilde{R}_e = R_e = \sigma^2$ in
(\ref{eq:Reinfexplicit}) - (\ref{eq:Pinfexplicit}) at $a=1$. Thus,
(20) has a value of zero at $a=1$, and Theorem
\ref{theo:errorexponentNPscalarstate} holds for $0\le a \le 1$.

\vspace{0.5em} Now we show that (\ref{eq:errorexponentscalar}) is
equivalent to  the spectral domain result
(\ref{eq:errorexponentspectralnewderiv}) using spectral
factorization. From the spectral domain form
(\ref{eq:errorexponentspectralnewderiv}) we have
\begin{eqnarray}
K &=& \frac{1}{4\pi} \int_0^{2\pi} \log\frac{S_y^{(1)}(\omega)}{
\sigma^2}d\omega + \frac{1}{4\pi}
\int_0^{2\pi}\frac{\sigma^2}{S_y^{(1)}(\omega)}d\omega
 - \frac{1}{2},\nn\\
&=& \frac{1}{4\pi} \int_0^{2\pi} \log S_y^{(1)}(\omega)d\omega +
\frac{1}{4\pi}
\int_0^{2\pi}\frac{\sigma^2}{S_y^{(1)}(\omega)}d\omega
-\frac{1}{2}\log \sigma^2 - \frac{1}{2}.
\label{eq:ProofKspectEquiv1}
\end{eqnarray}
First, consider the first term on the RHS of
(\ref{eq:ProofKspectEquiv1}).  The argument of the logarithm is
the power spectral density of the observation sequence $\{y_i\}$
under $H_1$. From Wiener filtering theory, the canonical spectral
factorization for $S_y^{(1)}(z)$ is given by
(\cite{Kailath&Sayed&Hassibi:book}, p.275)
\begin{eqnarray}
S_y^{(1)}(z)&=&L(z)R_e L^*(z^{-*}), \label{eq:spectralfactor}
\end{eqnarray}
where  $L^{-1}(z)$ is the whitening filter.
 Hence, we have
\begin{eqnarray*}
&&\frac{1}{4\pi} \int_0^{2\pi} \log S_y^{(1)}(\omega) d\omega\\
&=& \frac{1}{4\pi} \int_0^{2\pi} \log (  L(e^{j\omega})R_e L^*( e^{j\omega})   )d\omega,\\
&=&  \frac{1}{4\pi} \int_0^{2\pi} ( \log R_e + \log L(e^{j\omega}) + \log L^*( e^{j\omega})    ) d\omega,\\
&=& \frac{1}{2}\log R_e,
\end{eqnarray*}
where the last step follows from the cancellation of  two other
terms in para-Hermitian conjugacy.  Now, consider the second term
on the RHS of  (\ref{eq:ProofKspectEquiv1}).  From
(\ref{eq:spectralfactor}), we have
\begin{equation}
 \frac{\sigma^2}{S_y^{(1)}(z)}=  \frac{\sigma^2 L^{-1}(z)( L^*(z^{-*}))^{-1}}{R_e},
\end{equation}
which is the spectral density of the innovations process under
$H_0$ divided by $R_e$, since $\{y_i\}$ is an i.i.d. sequence with
variance $\sigma^2$ under $H_0$ and $L^{-1}(z)$ is the whitening
filter. Since the variance of a stationary process is given by the
autocovariance function $r(l)$ setting $l=0$, we have, by the
definition of $\tilde{R}_e$,
\begin{equation}   \label{eq:tildeReinftyspectral}
\tilde{R}_e= r(0)= \frac{1}{2\pi} \int_0^{2\pi} \sigma^2 [L^{-1}(z)( L^*(z^{-*}))^{-1}]_{z=e^{j\omega}}  d\omega
\end{equation}
since the spectral density is the Fourier transform of the
autocovariance function.  (Eq. (\ref{eq:tReinfexplicit}) is an
explicit formula for (\ref{eq:tildeReinftyspectral}).)   Hence, we
have
\begin{equation}
\frac{1}{4\pi} \int_0^{2\pi}
\frac{\sigma^2}{S_y^{(1)}(\omega)}d\omega = \frac{1}{2}
\frac{\tilde{R}_e}{R_e},
\end{equation}
and (\ref{eq:ProofKspectEquiv1}) is given by
\begin{equation}
 \frac{1}{2} \log R_e
+\frac{1}{2}\frac{\tilde{R}_e}{R_e}  -\frac{1}{2}\log \sigma^2-
\frac{1}{2}=  \frac{1}{2} \log \frac{R_e}{\sigma^2}
+\frac{1}{2}\frac{\tilde{R}_e}{R_e}  - \frac{1}{2} ,
\end{equation}
which is the error exponent in Theorem
\ref{theo:errorexponentNPscalarstate} that we derived using the
innovations approach. \hfill{$\blacksquare$}

\vspace{1em}
\begin{lemma} \label{lemma:derivativeofeta}
The partial derivative  of the error exponent with respect to the
correlation coefficient $a$ is given by
\begin{equation}
\frac{\partial K}{\partial a} = \frac{\Gamma (b-a)}{ r_e(1-b^2)}
\left(
\frac{1}{1-ab} - \frac{2(1-ab)}{r_e(1-b^2)^2}
\right)
\end{equation}
for a fixed SNR $\Gamma$,  where $b= a/r_e$ and $r_e=R_e/\sigma^2$.
\end{lemma}

\vspace{1em} It is easily seen that the partial derivative
$\frac{\partial K}{\partial a}$ is a continuous function of $a$
for $0 \le a < 1$ since $r_e$ is a continuous function of $a$ from
(\ref{eq:Reinfexplicit}) and (\ref{eq:Pinfexplicit}).

\vspace{1em}
{\em Proof of Lemma~\ref{lemma:derivativeofeta}}

We use the spectral domain form for the error exponent.
\begin{eqnarray}
K &=&
-\frac{1}{4\pi} \int_0^{2\pi} \log\frac{\sigma^2}{\sigma^2+S_s(\omega)}d\omega
+ \frac{1}{4\pi} \int_0^{2\pi}\frac{\sigma^2}{\sigma^2+S_s(\omega)}d\omega
 - \frac{1}{2},\nn\\
&=& -\frac{1}{4\pi} \int_0^{2\pi} \log\frac{1}{1+\Gamma\tilde{S}_s(\omega)}d\omega
+ \frac{1}{4\pi} \int_0^{2\pi}\frac{1}{1+\Gamma\tilde{S}_s(\omega)}d\omega
 - \frac{1}{2},\label{eq:etaspectralprooflemdev}
\end{eqnarray}
where
\begin{equation}   \label{eq:tildeSsomega}
S_s(\omega)= \frac{\Pi_0(1-a^2)}{1-2 a \cos \omega +a^2}, ~~~\tilde{S}_s(\omega)=S_s(\omega)/\Pi_0, ~~~\Gamma=\frac{\Pi_0}{\sigma^2}.
\end{equation}
The  spectral density of the observation sequence $\{y_i\}$  is
given by
\begin{eqnarray}
S_y^{(1)}(z)&=& \sigma^2+S_s(z)=\sigma^2 \left(1 +
\frac{Q/\sigma^2}{(1-az^{-1})(1-az)}\right),
\end{eqnarray}
where $Q=\Pi_0(1-a^2)$, and its canonical spectral factorization is given by (\cite{Kailath&Sayed&Hassibi:book}, p.242)
\begin{equation}  \label{eq:prooflemma4_syz}
S_y^{(1)}(z)= \sigma^2 L(z)r_e L^*(z^{-*})= \sigma^2r_e
\frac{1-bz}{1-az}\frac{1-bz^{-1}}{1-az^{-1}},
\end{equation}
where $b=a/r_e$ ($ |a| < 1 $ and $|b| < 1$) and
\begin{equation} \label{eq:smallreinproof}
r_e =\frac{\sqrt{[1+a^2+Q/\sigma^2]^2 -4a^2}+1+a^2+Q/\sigma^2}{2} =\frac{R_e}{\sigma^2}.
\end{equation}
The partial derivative of $K$ with respect to $a$ is given by
\begin{equation} \label{eq:etaderivative1}
\frac{\partial K}{\partial a}
= \frac{1}{4\pi} \int_0^{2\pi} \frac{\Gamma\tilde{S}_s^\prime(\omega)}{1+\Gamma\tilde{S}_s(\omega)}d\omega
   -\frac{1}{4\pi} \int_0^{2\pi} \frac{ \Gamma\tilde{S}_s^\prime(\omega)}{(1+\Gamma\tilde{S}_s(\omega))^2}d\omega ,
\end{equation}
where
\begin{equation}
\tilde{S}_s^\prime(\omega)= \frac{\partial \tilde{S}_s(\omega)}{\partial a}
= \frac{2[(1+a^2)\cos \omega -2a]}{(1-2a\cos \omega + a^2)^2}.
\end{equation}
Consider the first term on the RHS of  (\ref{eq:etaderivative1}).
Using the canonical spectral decomposition
(\ref{eq:prooflemma4_syz}), we have
\begin{eqnarray}
\frac{1}{4\pi} \int_0^{2\pi} \frac{\Gamma\tilde{S}_s^\prime(\omega)}{1+\Gamma\tilde{S}_s(\omega)}d\omega
&=&\frac{1}{4\pi} \int_0^{2\pi}   \frac{\frac{2\Gamma[(1+a^2)\cos \omega -2a]}{(1-2a\cos \omega + a^2)^2}   }{~~~~r_e \frac{1-be^{j\omega}}{1-ae^{j\omega}}\frac{1-be^{-j\omega}}{1-ae^{-j\omega}}~~~~ }  d\omega,  \nn \\
&=&\frac{1}{4\pi} \oint  \frac{\Gamma[(1+a^2)(z+z^{-1})-4a]}{r_e(1-az)(1-az^{-1})(1-bz)(1-bz^{-1})} \frac{dz}{jz} ,  \nn \\
&=&\frac{1}{4\pi j} \oint  \frac{\Gamma[(1+a^2)(z^2+1)-4az]}{r_e(1-az)(z-a)(1-bz)(z-b)} dz,  \nn \\
&=& \frac{1}{4\pi j} \frac{\Gamma}{r_e} 2\pi j \sum_{|z| < 1} \mbox{Residues of integrand},\nn \\
&=& \frac{\Gamma}{2r_e}  \left( \frac{1-a^2}{(1-ab)(a-b)} + \frac{(1+a^2)(1+b^2)-4ab}{(1-ab)(b-a)(1-b^2)} \right),\nn\\
&=& \frac{\Gamma}{r_e}\frac{(b-a)}{(1-ab)(1-b^2)},
\end{eqnarray}
where we have substituted $z=e^{j\omega}$, and used the residue
theorem. The second term on the RHS of  (\ref{eq:etaderivative1})
is similarly obtained:
\begin{eqnarray}
\frac{1}{4\pi} \int_0^{2\pi} \frac{ \Gamma\tilde{S}_s^\prime(\omega)}{(1+\Gamma\tilde{S}_s(\omega))^2}d\omega
&=& \frac{1}{4\pi}  \oint \frac{\Gamma[(1+a^2)(z+z^{-1})-4a]}{r_e^2(1-bz)^2(1-bz^{-1})^2} \frac{dz}{jz},\nn\\
&=& \frac{\Gamma}{4\pi r_e^2 j}   \oint \frac{(1+a^2)(z^2+1)-4az}{(1-bz)^2(z-b)^2} dz,\nn\\
&=& \frac{\Gamma}{4\pi r_e^2 j} 2\pi j  \mbox{Res} (z=b),\nn
\end{eqnarray}
where
\begin{equation}
 \mbox{Res} (z=b)=    \left[\frac{\partial}{\partial z}\frac{(1+a^2)(z^2+1)-4az}{(1-bz)^2}\right]_{z=b}= \frac{4(b-a)(1-ab)}{(1-b^2)^3}.
\end{equation}
Hence, $K^\prime$ is given by
\begin{equation} \label{eq:etaderivativeproof}
\frac{\partial K}{\partial a} =
\frac{\Gamma(b-a)}{r_e(1-b^2)}\left( \frac{1}{1-ab} -
\frac{2(1-ab)}{r_e(1-b^2)^2} \right).
\end{equation}
\hfill{$\blacksquare$}

\vspace{2em}
{\em Proof of Theorem~\ref{theo:etavscorrelation}}

First, the continuity of $K$ in (\ref{eq:errorexponentscalar}) is
straightforward as a function of $R_e$ and $\tilde{R}_e$ since
$R_e \ge \sigma^2$, and the continuity of $R_e$ and $\tilde{R}_e$
is also trivial as a function of $a$ and $P$ from
(\ref{eq:Reinfexplicit}) and (\ref{eq:tReinfexplicit}). Thus, we
need only to show the continuity of $P$ as a function of $a$,
i.e., the nonnegativity of the argument of the square root in
(\ref{eq:Pinfexplicit}). The argument  can be rewritten as
\begin{equation}
[\sigma^2(1-a^2)-Q]^2+ 4\sigma^2
Q=\sigma^4(1-a^2)[(1-a^2)(1-\Gamma)^2+4\Gamma],
\end{equation}
which is nonnegative if either $\Gamma=1$ or
$\frac{-4\Gamma}{(1-\Gamma)^2} \le \min_{a \in [0,1]} (1-a^2)=0$
if $\Gamma\ne 1$. Thus, $K$ is a continuous function of $a$
~$(0\le a \le 1)$ for any SNR $\Gamma \ge 0$.

{\em (i) SNR $\ge 1$:}

Since $b=a/r_e$ in (\ref{eq:etaderivativeproof}), we have
\begin{eqnarray}
\frac{\partial K}{\partial a} &=&\frac{\Gamma(b-a)}{r_e(1-b^2)}\left( \frac{1}{1-ab} - \frac{2b(1-ab)}{a(1-b^2)^2} \right),\nn\\
&=&\frac{\Gamma(b-a)}{r_e(1-b^2)}\frac{[a(1-b^2)^2-2b(1-ab)^2)]}{a(1-ab)(1-b^2)^2}, \nn \\
&=& \frac{\Gamma(b-a)}{r_e(1-b^2)}\frac{[a(1+b^2)^2-2b(1+a^2b^2)]}{a(1-ab)(1-b^2)^2}, \nn \\
&=& \frac{\Gamma(b-a)[br_e(1+b^2)^2-2b(1+a^2b^2)]}{r_e a(1-ab)(1-b^2)^3}, \nn \\
&=& \frac{\Gamma (b-a)b r_e^{-1} [(1+\tilde{Q}+a^2)^2-2r_e(1+a^4r_e^{-2})]}{r_e a(1-ab)(1-b^2)^3}, \label{eq:etahighsnr1}
\end{eqnarray}
where  $\tilde{Q}\defeq\Gamma (1-a^2)$ and we have used the relation
$r_e(1+b^2) = 1 + \tilde{Q}+a^2$ in the canonical spectral
factorization. (See \cite{Kailath&Sayed&Hassibi:book} p.242.) We
also have the relation
\begin{equation}
r_e + \frac{a^2}{r_e} = 1+\tilde{Q}+a^2,
\end{equation}
which implies
\begin{equation}
r_e(1+a^4r_e^{-2}) \le 1+\tilde{Q}+a^2,
\end{equation}
since $0 \le a < 1$. Hence, for  the last term in the numerator of (\ref{eq:etahighsnr1}) we have
\begin{equation}  \label{eq:numerator82}
(1+\tilde{Q}+a^2)^2-2r_e(1+a^4r_e^{-2}) \ge (1+\tilde{Q}+a^2)^2 - 2(1+\tilde{Q}+a^2).
\end{equation}
The RHS of (\ref{eq:numerator82}) is positive for $1+\tilde{Q}+a^2
= 1 + a^2+ \Gamma (1-a^2) \ge 2$, which reduces to the condition
$\Gamma \ge 1$.  Since
\begin{equation}
r_e = R_e/\sigma^2 = 1 + P/\sigma^2 > 1 ~~\mbox{for}~~ 0 \le a < 1,
\end{equation}
we have
\begin{equation}
b-a= (r_e^{-1} -1 ) a < 0.
\end{equation}
Hence,  $\frac{\partial K}{\partial a} \le  0$ for $0\le a < 1$
and $\Gamma \ge 1$, and $K$ is monotonically decreasing as $a
\uparrow 1$ for $\Gamma \ge 1$.

\vspace{1em}
{\em (ii) SNR $<1$:}

For a given $\Gamma$, denote the last term in the numerator of
(\ref{eq:etahighsnr1}) by
\begin{equation}
f_\Gamma (a) \defeq (1+\tilde{Q}+a^2)^2-2r_e(1+a^4r_e^{-2}).
\end{equation}
 Then, we can write
\begin{eqnarray}
f_\Gamma (a=0) &=& (1+ \Gamma)^2 -2(1+\Gamma)=\Gamma^2 -1,
\end{eqnarray}
since $\tilde{Q}=\Gamma(1-a^2)$ and $r_e =1+\Gamma$ for $a=0$ from
(\ref{eq:smallreinproof}). We have $f_\Gamma (a=0) < 0$ for
$\Gamma <1$ and $\frac{\partial K}{\partial a} > 0$
 from $(\ref{eq:etahighsnr1})$  since $b-a <0$. Hence, $K$ increases as $a$ increases
  in the neighborhood of $a=0$ with $K(a=0)= D(\Nc(0,1)||\Nc(0,1+\Gamma) > 0$ if  $0 < \Gamma < 1$.
   However, $K \downarrow 0$ as $a$ approaches  one since $P_M \sim \Theta(\frac{1}{\sqrt{n}})$ at $a=1$.
    Hence, $K$ achieves a maximum at nonzero $a$  for SNR $\Gamma <1$ since $K$ is a continuous function of $a$, and the value of $a$ achieving the maximum is given by  $f_\Gamma(a) = 0$ since $\frac{\partial K}{\partial a}$ is also continuous with $a$.

As SNR $\Gamma \downarrow 0$, we have
\begin{equation}
\tilde{Q}=\Gamma (1-a^2) \downarrow 0 ~~\mbox{and}~~ r_e = 1+ P/\sigma^2 \downarrow  1.
\end{equation}
The last term in the numerator of (\ref{eq:etahighsnr1}) is given by
\begin{eqnarray}
(1+\tilde{Q}+a^2)^2-2r_e(1+a^4r_e^{-2}) &\rightarrow& (1+a^2)^2 - 2(1+a^4),\nn\\
&=& - (1-a^2)^2 < 0,
\end{eqnarray}
for $0 \le  a < 1$ as $\Gamma \downarrow 0$.  Hence, for any
$\delta
> 0$, there exists $\Gamma_0$ small enough such that for all
$\Gamma < \Gamma_0(\delta)$, $(1+ \tilde{Q} +a^2)^2- 2r_e(1 +
a^4r_e^{-2}) \le -(1-a^2)^2 + \delta$. This guarantees that for
$\Gamma < \Gamma_0(\delta), \partial K/\partial a >0$ for $0 < a <
\sqrt{1-\sqrt{\delta}}$, and $a^* \ge \sqrt{1-\sqrt{\delta}}$.
\hfill $\blacksquare$

\vspace{2em}
{\em Proof of Theorem~\ref{theo:etavsSNR}}

Let $s=1 + \Gamma\tilde{S}_s(\omega)$ where $\tilde{S}_s(\omega)$ is
given by (\ref{eq:tildeSsomega}). Then, from
(\ref{eq:etaspectralprooflemdev}), the partial derivative of $K$
w.r.t. $\Gamma$ is given by
\begin{equation}
\frac{\partial K}{\partial \Gamma} = \frac{1}{2\pi} \int_0^{2\pi}
\frac{\partial }{\partial s}\left(-\frac{1}{2} \log \frac{1}{s}
+ \frac{1}{2s} -\frac{1}{2}\right)   \frac{\partial s}{\partial \Gamma } d\omega,
\end{equation}
where
\begin{equation}
 \frac{\partial }{\partial s}
\left(-\frac{1}{2}\log\frac{1}{s} +\frac{1}{2}\frac{1}{s}
-\frac{1}{2}\right) =
\frac{1}{2}\frac{s-1}{s^2}=\frac{1}{2}\frac{\Gamma
\tilde{S}_s(\omega)}{s^2} > 0,
\end{equation}
and
\begin{equation}
\frac{\partial s}{\partial \Gamma } = \tilde{S}_s(\omega)=\frac{1-a^2}{1-2a\cos(\omega)+a^2} > 0
\end{equation}
for $0\le a <1$.  Hence,
\begin{equation}
\frac{\partial K}{\partial \Gamma} > 0,
\end{equation}
and the error exponent $K$ increases monotonically as SNR increases
for a given $a$ ($ 0 \le a < 1$).

At high SNR, we have
\begin{eqnarray*}
P &\sim& Q,\\
R_e &\sim& Q+\sigma^2,\\
\tilde{R}_e &\sim& \sigma^2(1+a^2).
\end{eqnarray*}
Hence, from (\ref{eq:errorexponentscalar}), the error exponent is
given at high SNR by
\begin{eqnarray}
K &\sim& \frac{1}{2}\log \frac{Q+\sigma^2}{\sigma^2} + \frac{1}{2}
\frac{\sigma^2(1+a^2)}{Q+\sigma^2},\nn \\
&=& \frac{1}{2} \log \frac{Q+\sigma^2}{\sigma^2} + \frac{1}{2}
\frac{1+a^2}{\frac{Q+\sigma^2}{\sigma^2}},
    ~~~~ Q = \Pi_0 (1-a^2),\nn \\
&=& \frac{1}{2} \log \left( 1+\frac{\Pi_0}{\sigma^2}(1-a^2)\right)
+ \frac{1}{2} \frac{1+a^2}{1+\frac{\Pi_0}{\sigma^2}(1-a^2)}.
\label{eq:etahighSNRfinal}
\end{eqnarray}
Since the first term is dominant at high SNR, the theorem follows.
\hfill $\blacksquare$

\vspace{1em}
{\em Proof of Theorem~\ref{theo:errorexponentNPvector}}

 Since the  error exponent is given by the
asymptotic Kullback-Leibler rate (\ref{eq:asymKLrate}) and its
representation by innovations for the vector
 case is the same as (\ref{eq:ProofKInnovp0}) and (\ref{eq:ProofKInnovfirstterm}).
 We need only to calculate $R_e$ and $\tilde{R}_e$ for the vector model.

The steady-state variance $R_e$ for the innovations under $H_1$ is
given by the conventional result of the state-space model
\begin{equation} R_e = \hbf^T \Pbf \hbf + \sigma^2, \end{equation} and
$\Pbf$ is the unique Hermitian solution of the discrete-time
Riccati equation \begin{equation}  \label{eq:proof1DTRiccati} \Pbf
= \Abf \Pbf \Abf^T + \Bbf\Qbf\Bbf^T - \frac{\Abf \Pbf \hbf \hbf^T
\Pbf \Abf^T}{ \hbf^T \Pbf \hbf+\sigma^2}, \end{equation} such that
$\Abf - \Kbf_p \hbf^T$ is stable (the existence of the solution is
guaranteed since $\Abf$ is stable, $\mbox{diag}(\Qbf, \sigma^2)
\ge 0$, and $S_y^{(1)}(\omega) > 0$ due to the additive noise. See
\cite{Kailath&Sayed&Hassibi:book} p. 277.), where
\begin{equation} \Kbf_{p} = \Abf\Pbf \hbf R_e^{-1}.
\end{equation}
For $\tilde{R}_e$, we again represent $e_i$ as a linear
combination of $y_1, y_2,\ldots, y_i$, and $e_i$ is given by
\begin{eqnarray}
e_i &=& ~y_i -\hbf^T \Kbf_{p,i-1}y_{i-1} -\hbf^T(\Abf-\Kbf_{p,i-1}\hbf^T)\Kbf_{p,i-2}y_{i-2} \nn\\
&& ~~~-\hbf^T(\Abf-\Kbf_{p,i-1}\hbf^T)(\Abf-\Kbf_{p,i-2}\hbf^T)\Kbf_{p,i-3}y_{i-3}\nn\\
    && ~~~-\hbf^T(\Abf-\Kbf_{p,i-1}\hbf^T)(\Abf-\Kbf_{p,i-2}\hbf^T)(\Abf-\Kbf_{p,i-3}\hbf^T)\Kbf_{p,i-4}y_{i-4} - \cdots
\end{eqnarray}
where $\Kbf_{p,i}$ is the Kalman prediction gain given by
$\Kbf_{p,i}=\Abf \Pbf_i \hbf /(\hbf^T \Pbf_i \hbf + \sigma^2)$ with
the one-step prediction error covariance matrix
$\Pbf_i$\cite{Kailath&Sayed&Hassibi:book}. Since the Kalman filter
converges  asymptotically to the time-invariant recursive Wiener
filter when $\Abf$ is stable, we have asymptotically
\begin{eqnarray}
e_i &=& ~y_i -\hbf^T \Kbf_{p}y_{i-1} -\hbf^T(\Abf-\Kbf_{p}\hbf^T)\Kbf_{p}y_{i-2} \nn\\
&& ~~~-\hbf^T(\Abf-\Kbf_{p}\hbf^T)(\Abf-\Kbf_{p}\hbf^T)\Kbf_{p}y_{i-3}\nn\\
    && ~~~-\hbf^T(\Abf-\Kbf_{p}\hbf^T)(\Abf-\Kbf_{p}\hbf^T)(\Abf-\Kbf_{p}\hbf^T)\Kbf_{p}y_{i-4} - \cdots
\end{eqnarray}
where $\Kbf_{p}$ is the steady-state Kalman prediction gain. Thus,
the innovation sequence becomes the output of a stable recursive
filter driven by the i.i.d. sequence $\{y_i\}$ under $H_0$ as in the
scalar case, and the  ergodic theorem holds for
$\frac{1}{n}\sum_{i=1}^n e_i^2$.
\begin{eqnarray}
\tilde{R}_e &=& \lim_{i\rightarrow \infty} \Ebb_0 \{e_i^2\},\nn\\
&=& \sigma^2+\sigma^2 \hbf^T \left( \sum_{k=0}^\infty (\Abf-\Kbf_{p}\hbf^T)^k \Kbf_{p}\Kbf_{p}^T [(\Abf-\Kbf_{p}\hbf^T)^k]^T   \right) \hbf.
\end{eqnarray}
Let $\tilde{\Pbf}$ be defined as
\begin{equation}
\tilde{\Pbf} \defeq  \sum_{k=0}^\infty (\Abf-\Kbf_{p}\hbf^T)^k \Kbf_{p}\Kbf_{p}^T [(\Abf-\Kbf_{p}\hbf^T)^k]^T.
\end{equation}
$\tilde{\Pbf}$ is finite since  $\Abf - \Kbf_{p} \hbf^T$ is stable by the property of the solution of (\ref{eq:proof1DTRiccati}), and is given by the unique solution of the following Lyapunov equation
\begin{equation} \label{eq:LyapunovProof}
\tilde{\Pbf} -  (\Abf-\Kbf_{p}\hbf^T) \tilde{\Pbf}  (\Abf-\Kbf_{p}\hbf^T)^T = \Kbf_{p}\Kbf_{p}^T.
\end{equation}
(Since $\Abf - \Kbf_{p} \hbf^T$ is stable  and $\Kbf_{p}\Kbf_{p}^T
\ge 0$, there exists a unique, Hermitian, and positive
semi-definite solution $\tilde{\Pbf}$ of
(\ref{eq:LyapunovProof})\cite{Kailath&Sayed&Hassibi:book}.) The
spectrum for the vector model is given by
(\ref{eq:spectraldensityvector})
\cite{Kailath&Sayed&Hassibi:book}.

\hfill{$\blacksquare$}



\bibliographystyle{ieeetr}



\end{document}